# Quantum Spin Liquids Unveil the Genuine Mott State

A. Pustogow[1]*, M. Bories[1], A. Löhle[1], R. Rösslhuber[1], E. Zhukova[1,2], B. Gorshunov[1,2], S. Tomić[3], J.A. Schlueter[4,5], R. Hübner[1,6], T. Hiramatsu[7], Y. Yoshida[7], G. Saito[7,8], R. Kato[9], T.-H. Lee[10], V. Dobrosavljević[10], S. Fratini[11] and M. Dressel[1]

1. 1. Physikalisches Institut, Universität Stuttgart, Pfaffenwaldring 57, D-70569 Stuttgart Germany
2. Moscow Institute of Physics and Technology (State University), 141700, Dolgoprudny, Moscow Region, Russia
3. Institut za fiziku, P.O.Box 304, HR-10001 Zagreb, Croatia
4. Division of Materials Research, National Science Foundation, Arlington, VA 22230, USA
5. Materials Science Division, Argonne National Laboratory, Argonne, IL 60439, USA
6. Inst. Functional Matter and Quantum Techn., Universität Stuttgart, Pfaffenwaldring 57, D-70569 Stuttgart, Germany
7. Faculty of Agriculture, Meijo University, Nagoya 468-8502, Japan
8. Toyota Physical and Chemical Research Institute, Nagakute 480-1192, Japan
9. Condensed Molecular Materials Laboratory, RIKEN, 2-1 Hirosawa, Wako-shi, Saitama 351-0198, Japan
10. Department of Physics and National High Magnetic Field Laboratory, Florida State University, Tallahassee, Florida 32306, USA
11. Institut Néel - CNRS and Université Grenoble Alpes, 38042 Grenoble Cedex 9, France

**The Widom line identifies the locus in the phase diagram where a supercritical gas crosses over from gas-like to a more liquid-like behavior. A similar transition exists in correlated electron liquids, where the interplay of Coulomb repulsion, bandwidth and temperature triggers between the Mott insulating state and an incoherent conduction regime. Here we explore the electrodynamic response of three organic quantum spin liquids with different degrees of effective correlation, where the absence of magnetic order enables unique insight into the nature of the genuine Mott state down to the most relevant low-temperature region. Combining optical spectroscopy with pressure-dependent dc transport and theoretical calculations, we succeeded to construct a phase diagram valid for *all* Mott insulators on a quantitative scale. In the vicinity of the low-temperature phase boundary, we discover metallic fluctuations within the Mott gap, exhibiting enhanced absorption upon cooling that is not present in antiferromagnetic Mott insulators. Our findings reveal the phase coexistence region and Pomeranchuk-like anomaly of the Mott transition, previously predicted but never observed.**

The Mott metal-insulator transition (MIT) stands out among the key unresolved phenomena in interacting electron systems [1]. Sufficiently strong Coulomb repulsion is able to localize the charge carriers accompanied by a splitting of the nominally half-filled band into the lower (LHB) and upper (UHB) Hubbard bands separated by the Mott gap (Fig. 1b). When temperature becomes the dominant energy scale, i.e. comparable to the bandwidth $W$ and Coulomb interaction $U$, the Hubbard bands are strongly blurred by thermal broadening resulting in an incoherent conduction regime (Fig. 1d). While the Mott transition is a well-defined, first-order MIT below the critical endpoint, it shows analogy to the liquid-gas crossover in the supercritical high-temperature region [2,3], recently associated with quantum critical scaling [4]. In this part of the phase diagram, the quantum Widom line (QWL) defines the boundary of the fully gapped Mott state [5]. A similar description applies to the pseudogap in doped Mott insulators, which underpins the decisive role of Mott physics for high-temperature superconductivity in cuprates [6].

At low temperatures ($k_B T < 10^{-2}\,W$) magnetic instabilities typically mask the Mott MIT; the antiferromagnetic ground state dominates the low-energy excitations and, due to the Clausius-Clapeyron relation $dp/dT=\Delta S/\Delta V$, the small entropy implies a negative slope of the phase transition (Fig. 1f). To circumvent this problem, we study Mott insulators that are currently under scrutiny for their quantum spin liquid ground state (QSL) as a result of large geometrical frustration and disorder [7,8,9,10,11,12]. The absence of antiferromagnetism enables us to investigate the genuine Mott state down to $T \to 0$; due to the large spin degrees of freedom, the phase transition acquires a positive slope reminiscent of the Pomeranchuk-effect in $^3$He, where the solid (insulator) becomes liquid (metal) upon cooling (Fig. 1e). While from theory it is known that the phase diagram is controlled by the ratios $T/W$ and $U/W$ [13,5], where temperature and Coulomb repulsion are normalized to the bandwidth, experimental studies have not advanced beyond $T - p$ plots so far, inhibiting a quantitative comparison (i) between different compounds and (ii) with theory. For several organic charge transfer salts we succeeded to determine $U$ and $W$ from the optical spectra and combine our findings with pressure-dependent dc transport and dynamical mean-field theory calculations. By identifying the relevant scales that control the transition, our approach allows us to put *all* Mott insulators -- including non-frustrated compounds and transition metal oxides -- on one universal phase diagram in a quantitative way. Moreover, our optical data provide evidence for charge fluctuations close to the phase boundary on the insulating side, enhanced upon cooling due to the Pomeranchuk-like instability. Although theoretically predicted, this phase coexistence has never been observed before.

**Materials, Effective Models and Theoretical Studies**
For our investigation, we have selected three well-known organic Mott insulators $\beta'$-EtMe$_3$Sb [Pd(dmit)$_2$]$_2$ (abbreviated EtMe, dmit is 1,3-dithiole-2-thione-4,5-dithiolate), $\kappa$-(BEDT-TTF)$_2$ Ag$_2$(CN)$_3$ (abbreviated AgCN, BEDT-TTF denotes bis(ethylenedithio)tetrathiafulvalene) and $\kappa$-(BEDT-TTF)$_2$Cu$_2$(CN)$_3$ (abbreviated CuCN). These compounds form layered structure with dimers of organic molecules on a triangular lattice carrying one electron each, resulting in half-filled, quasi two-dimensional electronic systems, which are well described in terms of the single-band Hubbard model [10,14]. Charge transport is related to hopping between neighboring sites, proportional to the transfer integrals $t$ and $t'$ (Fig. 1a) and, thus, the bandwidth.

Panel d displays the prototypical phase diagram of a Mott insulator; the gap closes at the QWL (green data points determined by DMFT calculations, see *Methods*), which corresponds to a certain relation between $U$, $W$ and $T$. Since both scales (temperature $T/W$ and electronic correlations $U/W$) are expressed in units of $W$, the underlying physics is adequately described by only two parameters: the on-site Coulomb repulsion $U$ and the bandwidth $W$. As we will later see, the reversal of the QWL provides a fixed point to collapse the Widom lines of different compounds onto one universal curve. Strictly speaking, the bandwidth of the Hubbard bands $W$ is generally smaller than the bare bandwidth $W_0$ of the uncorrelated system. However, the effective values $U/W$ remain a good approximation of the intrinsic $U/W_0$ ratio even for strong correlations [15]. In order to determine the density of states (DOS) and optical conductivity spectra for different correlation strength $U/W$, we performed complementary DMFT calculations utilizing the continuous time quantum Monte Carlo (CTQMC) quantum impurity solver (*Methods* and *Supplementary information* S5, S7). The results affirm in particular that the maximum of the Mott-Hubbard band is proportional to the Coulomb repulsion $U$ (Fig. 1 **c**) and the $U/W$ ratio extracted from the band shape is closely related to the bare correlations $U/W_0$.

**Optical Studies**
Optical spectroscopy is the method of choice to experimentally determine the band parameters, as it directly maps the band shape by observing the transitions between the Hubbard bands, as illustrated in Fig. 1c. By measuring the infrared reflectivity of EtMe, AgCN and CuCN single crystals at different temperatures, we obtain the optical conductivity $\sigma_1(\omega)$ plotted in Fig. 2. Since the optical properties are rather isotropic within the highly-conducting plane, we focus on the polarization along the direction of the largest conductivity and present the data for the perpendicular axis in the *Supplementary information* S3. The most prominent feature is the Mott-Hubbard band centered around 2000 cm$^{-1}$, also seen in related materials [16,17,18,19]. Our CTQMC calculations reproduce the overall shape, asymmetry and intensity of the bands very well, establishing the dominant role of Mott physics in these compounds. The narrow peaks below 1500 cm$^{-1}$ correspond to molecular vibrations and lattice phonons [20,21], which can be clearly separated from the overall change of the Mott-Hubbard band in the contour plots.

As indicated in Fig. 2d-f, the maximum $\omega_{max} = \omega(\sigma_{max})$ and half-height $\omega_{1/2} = \omega(\sigma_{max}/2)$ frequencies, and thus the bandwidth determined as $W/\hbar = \omega_{max} - \omega_{1/2}$, change upon cooling. Due to the band asymmetry, we extract the bandwidth from the low-frequency wing of the Mott-Hubbard band for it is related to the intrinsic low-energy excitations of the ground state (Fig. 2g). The temperature-dependent band parameters, $U$ and $W$, and the corresponding degree of electronic correlations, $U/W$, are plotted in Fig. S5 of the *Supplementary information*. While the Mott-Hubbard bands of EtMe and AgCN are subject to narrowing upon cooling, a pronounced shift towards low frequencies and unusual broadening are observed for CuCN. In order to disentangle thermal effects and to focus on quantum Mott physics, we first analyze the data at the lowest temperatures; the temperature dependence of the low-frequency conductivity will be discussed in the context of Fig. 4.

From the spectroscopic data measured at $T = 5$ K, we conclude that electronic correlations are largest in EtMe despite the small $U$. Indeed, the small bandwidth in this material implies that (i) it is located most *left* in the $U/W$ phase diagram, (ii) the experimentally accessed temperatures (5 – 300 K) cover a broader vertical $T/W$ range and (iii) the electronic compressibility is largest because pressure has a larger effect on $U/W$. The two $\kappa$-compounds have larger energy scales and thus extend over smaller $T/W$ and $U/W$ ranges. Consequently, in a unified phase diagram the three materials are arranged in the order EtMe - AgCN - CuCN on the descending horizontal $U/W$ scale (Fig. 2h), consistent with *ab initio* density functional theory (DFT) and extended Hückel calculations (*Supplementary information* S1, Table S1) [22,23,24,25,11,8].

**Formation of the Mott Gap at the Quantum Widom Line**

A closer inspection of the temperature-dependent optical conductivity of EtMe (Fig. 3a and b), reveals a well-defined Mott gap that opens around 125 K and continuously grows up to 650 cm$^{-1}$ when the temperature is lowered to 5 K. This coincides well with the crossover temperature from Ref. [4] extrapolated to $p = 0$ establishing the QWL as the true boundary between the Mott state with a well-defined charge gap and the incoherent conduction regime in the supercritical region.

Panels c and d of Fig. 3 display the temperature-dependent dc conductivity and the corresponding transport gap $\Delta$ calculated from the logarithmic resistivity derivative d(ln$\rho$)/d(1/$T$) according to $\rho(T) \propto \exp(\Delta/k_B T)$, respectively. For all three compounds $\Delta(T)$ goes through an absolute maximum at a particular temperature, which coincides with the opening of the optical gap of EtMe (Fig. 3a) and can be associated with the QWL. There is similar agreement between Ref. [4] and our pressure-dependent dc transport results for CuCN revealing the QWL at 185, 135 and 123 K at $p = 0$, 2.17 and 2.38 kbar, respectively (Fig. S9). Although for AgCN there is a smaller feature at 90 K, Mott physics is reflected by the global maximum of $\Delta(T)$ implying that the QWL is crossed around 245 K. Therefore, the maximum of the transport gap allows us to identify the QWL even though no clear-cut optical gap is observed for CuCN and AgCN, probably due to the smaller $U/W$ ratio. We also notice that the size of the transport gap maximum scales with the Coulomb repulsion $U$ determined by optical spectroscopy.

**Low-Frequency Electrodynamic Response**

As the relevant energy scales in the phase diagram are of the order of 100 K or less, fingerprints of the ground state are expected in the low-frequency electrodynamic response, i.e. in the THz or far-infrared spectral ranges. A closer look on the temperature evolution of the optical conductivity well below $\omega_{max}$ (Fig. 4c) illustrates the insulating nature of the ground state of EtMe, where the sub-gap absorption diminishes upon cooling as thermal band broadening is reduced (cf. Fig. S6a). The integrated spectral weight, plotted in Fig. 4d,

$$SW(\omega_i) = \int_0^{\omega_i} \sigma_1(\omega) d\omega \xrightarrow{\hbar\omega_i < (U-W)} SW(\omega_i) \propto \frac{N}{m^*}, \qquad (1)$$

provides a robust measure of these low-energy excitations enabling a more quantitative analysis beyond dc transport. This quantity determines the absorption strength up to a given cutoff frequency $\omega_i$, chosen to be below the band edge ($U - W$) to probe the low-energy excitations only, and yields valuable information on correlation effects [28]. Here, $N$ is the charge carrier density and $m^*$ indicates the effective mass, renormalized by the band structure and electronic interactions.

While the overall spectral weight grows with decreasing electronic correlations (Fig. S7c), the temperature evolution $SW(\omega_i,T)$ plotted in Fig. 4 reveals an unexpected behavior. Although less pronounced than in EtMe, also AgCN behaves like a typical insulator, where the in-gap states are depleted upon cooling [29]. The opposite behavior is found in the case of CuCN, which exhibits a temperature dependence characteristic of metals (panel a, cf. Fig. S6b): the low-frequency optical conductivity increases upon lowering the temperature [30,31]. This is unexpected considering that no Drude peak is present, the hallmark of coherent transport, and that at zero frequency all compounds – including CuCN – are electrical insulators as determined from dc transport (Fig. 3c).

This apparent contradiction can be understood by the exceptional position of CuCN in the phase diagram: this compound is more proximate to the insulator-metal boundary than EtMe and AgCN (cf. Fig. 2h). Due to the Pomeranchuk-like anomaly, below the back-bending point of the QWL, the insulator-metal boundary gets closer to the ambient pressure position as the temperature is lowered (cf. Fig. 1e). Thus, we assign the strong non-thermal enhancement of the low-energy spectral weight below $T_{back}$ = 70 K to metallic fluctuations in the Mott state that appear upon entering the coexistence region close to the phase boundary, which has been predicted theoretically [13,5] but never observed. From the trend of our data we expect the anomalous slope of the Mott MIT to extend to the lowest measured temperature. Going back to Fig. 3d, the transport gap of CuCN gradually decreases and seems to vanish for $T \to 0$ consistent with a weakening of the insulating behavior. Therefore, the QWL not only marks a crossover in the conduction properties at elevated temperatures but is also manifest in the sub-gap region of the low-temperature optical absorption of CuCN.

If this was a fundamental property that occurs in general in the vicinity of the metallic state, it should also be identified in the optical data of the $\kappa$-(BEDT-TTF)$_2$Cu[N(CN)$_2$]Br$_x$Cl$_{1-x}$ series (detailed analysis in *Supplementary information* S6), where increasing Br content drives the bandwidth-tuned Mott MIT metallic and even superconducting [16,32,33,34,17]. Due to the substantially weaker frustration $t'/t$, the system orders antiferromagnetically at low temperatures, leading to a phase diagram like in Fig. 1f. Above $T_N$, however, these compounds behave similar as the frustrated Mott insulator CuCN and exhibit a comparable non-thermal enhancement of the SW corroborating our conclusions above. When antiferromagnetism is stabilized, a well-defined gap opens and the low-energy states are depleted; these intrinsic properties clearly distinguish magnetically ordered and QSL Mott states.

**Unified Phase Diagram**
Our comprehensive optical and transport results render possible the construction of the unified phase diagram of Fig. 5 that, on a quantitative scale, is valid for *all* frustrated Mott insulators. The universal quantum Widom line includes data points of EtMe and CuCN [4] and the maximum of dln($\rho$)/d$p$ calculated for AgCN [11]. On basis of our optical data, indicated by the left $T/W$ and bottom $U/W$ scales (cf. Fig. 2), we scaled the individual phase diagrams and merged them together (for details see *Supplementary information* S2). The horizontal and vertical bars on the top and right axes denote the pressure and temperature range (5 – 300 K) covered by the experimental data of each compound, respectively. In Fig. S12 we illustrate that, upon proper renormalization to the intrinsic energy scales, the inorganic Mott insulator (V$_{1-x}$Cr$_x$)$_2$O$_3$ fits well into the unified phase diagram, yet covering just a small range due to its large bandwidth.

The normalized pressure and temperature scales enable us to compare conveniently the experimental findings with results obtained from dynamical mean field theory (DMFT) calculations (Fig. 1d). The overall shape of the QWL, i.e. the characteristic back-bending at low temperatures, reminding of the Pomeranchuk effect, is consistently found in both experiment and theory. Even on a quantitative level, the $U/W$ and $T/W$ values match remarkably well, considering the present level of theory, i.e. the single-band Hubbard model as a valid effective low-energy description for the different compounds under study.

**Outlook**

By combining pressure-dependent transport data with optical conductivity, we can map the quantum Widom line up to an energy range never achievable in transition metal oxides, for they have a considerably larger bandwidth compared to the organic compounds studied here. Several thousands of Kelvins and pressures beyond the accessible experimental range are required to cover the entire phase diagram in Fig. 5 by similar experiments in vanadium oxides, for instance (cf. Fig. S12). In the overlapping region close to the reversal of the quantum Widom line, however, the behaviors agree rather well [35,36,37] indicating that the bandwidth is the proper scaling parameter and the underlying physics of genuine Mott insulators is universal. To further confirm the generality of our findings, it is desirable to extend these investigations to spin liquid compounds of different symmetry, e.g. kagome and other geometries; the intrinsic properties of the Mott MIT should prevail.

For transition metal compounds charge-carrier doping is the far more common way to go from the Mott insulating to the metallic regime, most prominent in the high-temperature superconducting cuprates. Despite enormous efforts devoted to this issue over the last 30 years, no general agreement was reached on a generic phase diagram that is in-depth understood comparable to the bandwidth-controlled Mott transition. Very recently, theoretical studies identified the pseudogap in hole-doped cuprates as a Widom line and ascribe the transport, dynamic and thermodynamic crossovers to a common origin linked to the intrinsic Mott state [6,38]. A similar rescaling approach might be appropriate to reveal the inherent relationship between Coulomb repulsion, doping and superconductivity in transition metal oxides. Moreover, evaluation of the scaling properties on the metallic side will provide important information on the emergence of strongly-interacting coherent quasiparticles from the interplay of $U$, $W$ and $T$.

We could trace the back-bending of the Mott insulator-metal boundary down to lowest temperatures and unequivocally identify the Pomeranchuk-like instability in the electrodynamics of $\kappa$-(BEDT-TTF)$_2$Cu$_2$(CN)$_3$; at finite frequencies, the compound becomes more metallic upon cooling although remaining an insulator in the dc-limit. This way we finally solved the long-standing dispute of excess conductivity and counterintuitive temperature dependence in the optical conductivity of this spin-liquid compound [30,39,31]. The anomalous slope of the Mott insulator-metal boundary seems to persist down to lowest temperatures. Such emerging metallic character may reflect the phase coexistence and spatially inhomogeneous structures caused by (weak) disorder in the vicinity of the phase transition. This interesting issue could be studied by systematically adding extrinsic disorder by heavy-ion irradiation [40,41,42]. The Anderson-Mott transition is subject of interest for decades [43], but here we provide a new possibility by exploring the dynamical properties of the fully frustrated case down to low temperatures and scaling the results in a quantitative manner.

## Methods

### Materials
Plate-like organic single crystals were grown during several months by air oxidation in acetone (EtMe) [23,44] and electrochemical oxidation (AgCN and CuCN) [45, 25] in RIKEN, Nagoya, Stuttgart and Argonne. The samples reach typical dimensions of $1 \times 1 \times 0.05$ mm³, $0.4 \times 0.3 \times 0.2$ mm³ and $1 \times 1 \times 0.2$ mm³ for $\beta'$-EtMe$_3$Sb[Pd(dmit)$_2$]$_2$, $\kappa$-(BEDT-TTF)$_2$Ag$_2$(CN)$_3$ and $\kappa$-(BEDT-TTF)$_2$Cu$_2$(CN)$_3$, respectively. After selection and cleaning, the crystals are measured without further treatment; in particular the optical reflectivity is probed on as-grown crystal surfaces.

### Optical Spectroscopy
Broadband optical spectra were recorded employing various Fourier-transform infrared spectrometers equipped with the corresponding sources, beam splitters, polarizers, windows, and detectors. The reflectivity was obtained by normalizing the signal off the sample to a gold mirror taking into account the tabulated frequency-dependent reflectivity of gold. To reach accurate absolute values and to account for diffraction effects in the far-infrared ($\omega/2\pi c < 700$ cm$^{-1}$), we first measured the temperature-dependent sample reflection normalized to a big mirror, followed by evaporation of a thin gold layer (approx. 300 nm thickness) on the sample and, subsequently, a second measurement of the covered sample at the same temperatures. Eventually, the reflection was calculated by dividing the sample spectra without and with gold, both of which were normalized to the big mirror to account for temporal fluctuations of the spectrometer. The low-frequency limit typically was 30 to 100 cm$^{-1}$ depending on the crystal size. The frequency resolution was set to 1 cm$^{-1}$.

Several optical cryostats enabled us to cover the range 5 K < $T$ < 300 K. Prior to cooling, the sample was aligned parallel to a gold mirror and the crystal axes were determined with an automated polarizer. Then the cryostat was evacuated to minimize thermal contact to the surroundings and avoid the formation of ice on the sample. During cooling, the typical pressure in the sample chamber was $10^{-6}$ - $10^{-5}$ mbar. The cooling rate was 1 K/min.

In addition to the Fourier-transform infrared reflection studies, temperature-dependent THz transmission was measured for $\beta'$-EtMe$_3$Sb[Pd(dmit)$_2$]$_2$ and $\kappa$-(BEDT-TTF)$_2$Cu$_2$(CN)$_3$, using coherent source (backward-wave oscillators) and pulsed time-domain spectrometers. Both spectrometers enable us to calculate the real and imaginary parts of the conductivity $\sigma_1(\omega) + i\sigma_2(\omega)$ directly without using Kramers-Kronig relations [46,28]. Thus, we could measure the single crystals of $\beta'$-EtMe$_3$Sb[Pd(dmit)$_2$]$_2$ and $\kappa$-(BEDT-TTF)$_2$Cu$_2$(CN)$_3$ down to 6 cm$^{-1}$; the low-frequency limit for $\kappa$-(BEDT-TTF)$_2$Ag$_2$(CN)$_3$ was around 100 cm$^{-1}$ due to the small crystal size. For all compounds the reflectivity was measured up to the visible (20000 cm$^{-1}$) by Fourier-transform spectroscopy. In addition, we determined the room temperature reflectivity of $\beta'$-EtMe$_3$Sb[Pd(dmit)$_2$]$_2$ up to 45000 cm$^{-1}$ by ellipsometric techniques.

### Data Processing
In order to calculate the frequency-dependent conductivity by the Kramers-Kronig relations, the optical reflectivity (see Figs. S2 and S3) obtained by the various methods and spectrometers was merged. In rare cases it was required to slightly shift the raw data (several percent mismatch can occur due to alignment in different setups). Then, the low- and high-frequency reflectivity was extrapolated to 0.001 and $10^6$ cm$^{-1}$, respectively. While generalized Hagen-Rubens behavior ($R = a - b\,\omega^n$ where $0 < n < 0.5$) was extrapolated at low frequencies, a $\omega^{-4}$ decay was applied towards high frequencies. For $\beta'$-EtMe$_3$Sb[Pd(dmit)$_2$]$_2$ and $\kappa$-(BEDT-TTF)$_2$Cu$_2$(CN)$_3$, the low-frequency optical conductivity resulting from the Kramers-Kronig calculation was replaced by the THz data. Most importantly, the optical conductivity discussed here is not affected by the extrapolation.

**Electrical Transport**
Temperature-dependent measurements of the dc resistivity were performed within the plane using the standard four-probe method with a typical current of 10 µA to avoid heating of the sample. The contacts were made by pasting 25 µm gold wires with a small amount of carbon paint directly on the crystal. The specimens were mounted on a sapphire plate for good thermal contact. Temperature-dependent measurements were conducted in a $^4$He exchange gas cryostat down to 2 K. The samples were slowly cooled down with approximately 0.1 K/min to avoid cracks, to ensure thermal equilibrium and to minimize effects associated with disorder.
Pressure-dependent experiments on $\kappa$-(BEDT-TTF)$_2$Cu$_2$(CN)$_3$ were performed in a CuBe oil pressure cell. The pressure was corrected for the temperature-dependent freezing of the medium. In Fig. 5 and S1 we incorporated the data of the pressure and temperature dependent resistivity measurements of Furukawa *et al.* and Shimizu *et al.* [4,11] on $\beta'$-EtMe$_3$Sb [Pd(dmit)$_2$]$_2$, $\kappa$-(BEDT-TTF)$_2$Ag$_2$(CN)$_3$, and $\kappa$-(BEDT-TTF)$_2$Cu$_2$(CN)$_3$. As sketched in Fig. 5 right insets, the inflection points of the resistivity versus pressure are taken as a crossing of the quantum Widom line at certain temperatures. It agrees well with the determination from our temperature-dependent transport studies presented in Fig. 3.

**Dynamical Mean Field Theory Calculations**
Dynamical Mean Field Theory (DMFT) [47] calculations were performed using a single-band Hubbard model at half-filling, and assuming a simple semi-circular density of states. We used the state-of-the-art Continuous Time Quantum Monte Carlo (CTQMC) quantum impurity solver, as first implemented by Haule [48]. The Monte Carlo step was set to $1\times10^9$ for precise sampling. In the insulating phase, the convergence was checked by requiring the local Green's function to satisfy the self-consistent criterion, $|G^{i+1}(i\omega_0)-G^i(i\omega_0)| \sim 10^{-4}$. This was typically achieved within 20 iterations, where $i$ represents the iterations and $\omega_0=\pi T$. We performed analytical continuation on the Matsubara frequency Green's function using Maximum Entropy Methods [49]. Then, the self-energy was calculated from the DMFT self-consistent condition:
$$\Sigma(\omega) = \omega + \mu - G^{-1}(\omega) - t^2 G(\omega). \qquad (2)$$
We set the half-bandwidth $D = \frac{1}{2}W_0 = 1$ as our energy unit. With the obtained self-energy, we calculate the optical conductivity using the Kubo formula [47]:
$$\sigma(\omega) = \sigma_0 \iint d\varepsilon d\nu \Phi(\varepsilon) A(\varepsilon,\nu) A(\varepsilon,\nu+\omega) \frac{f(\nu)-f(\nu+\omega)}{\omega}, \qquad (3)$$
where $A(\varepsilon,\nu)=-(1/\pi)\text{Im}(\varepsilon+\mu-\nu-\Sigma(\varepsilon))^{-1}$, $f$ is the Fermi function, $\Phi(\varepsilon)= \Phi(0)[1-(\varepsilon/D)^2]^{3/2}$, and $\sigma_0=2\pi e^2/\hbar$. Here, we set $\sigma_{IRM}=e^2\Phi(0)/\hbar D$ as our conductivity unit, where $\sigma_{IRM}$ is the Ioffe-Regel-Mott limit for distinguishing the metal and bad metallic behavior.

For evaluating the quantum Widom line (QWL) the DMFT calculation was performed on the triangular lattice to compare with the organic compounds in this report. To obtain a reliable QWL, the iterated perturbative theory (IPT) was applied as our impurity solver, which has been shown to have accurate results compared to numerically exact CTQMC but suffer from less numeric error, especially in triangular lattice systems. We employed the $\lambda$ analysis to locate the QWL [13,5], where $\lambda$ corresponds to the curvature of the free energy functional and shows a minimum at the QWL. More specifically, $\lambda$ can be obtained from the iterative solution of DMFT equations as described in Ref. [13,5],

$$\delta G^{(n+1)}(i\omega_n) - \delta G^{(n)}(i\omega_n) = e^{-n\lambda} \delta G^{(0)}(i\omega_n). \qquad (4)$$

where $\delta G^{(n)} = G^{(n)} - G_{DMFT}$ and $G_{DMFT}$ is the converged DMFT Green's function. Therefore, $\lambda$ corresponds to the convergence rate towards the DMFT solution.

# Figures

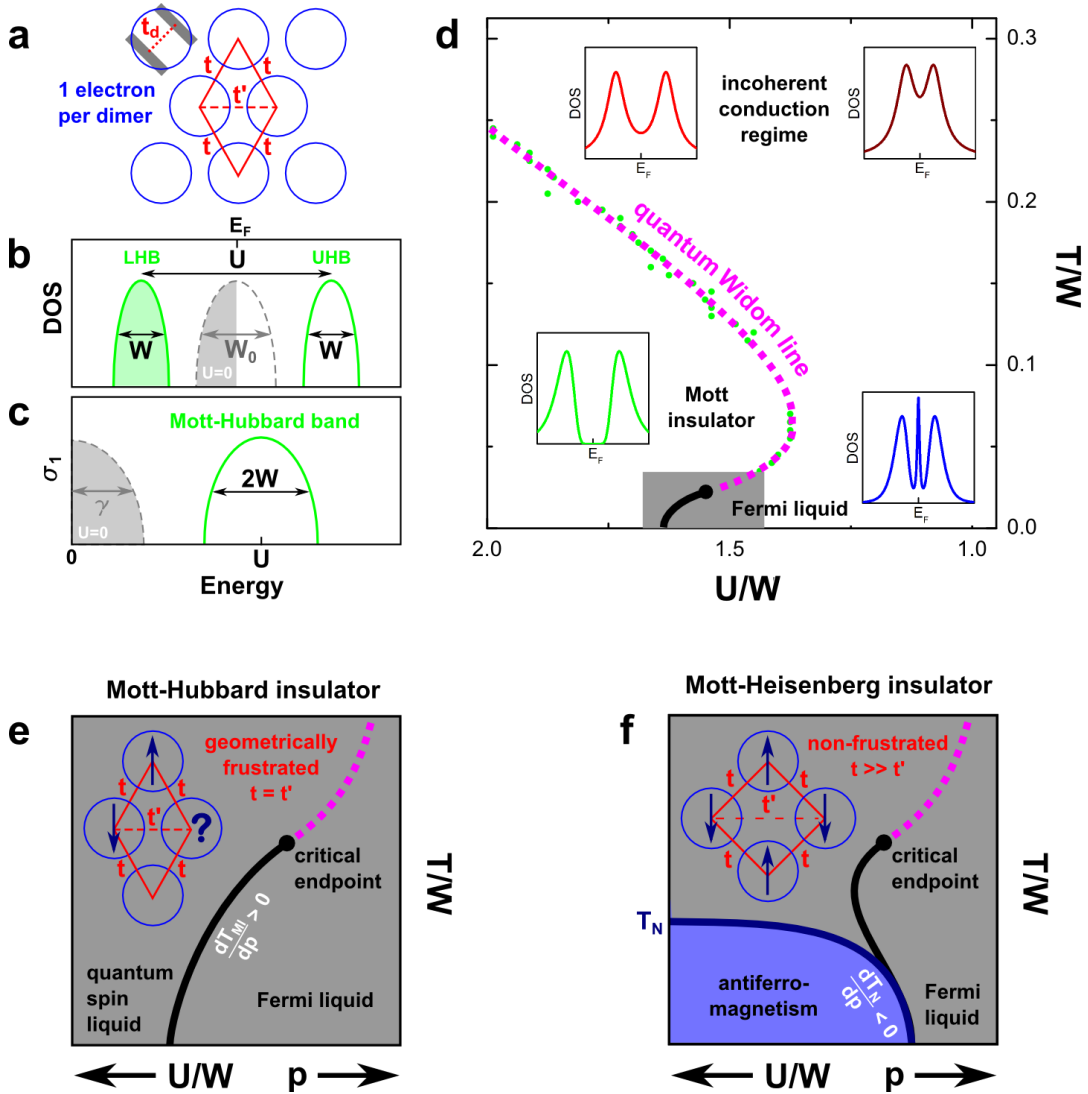

**Figure 1 Electronic excitations of the Mott insulator on a half-filled triangular lattice.**
**a**, The compounds under study have a layered structure with dimers (blue circles) of organic molecules (gray bars) arranged on a triangular lattice, where $t_d > t$ ($t'$) are the intra- and inter-dimer transfer integrals. **b**, Density of states (DOS) of the Mott insulator. The separation of the lower and upper Hubbard bands (green) is a measure of the Coulomb repulsion $U$ (*Supplementary information* S1). Although generally smaller, the width $W$ of the Hubbard bands remains a good measure of the bare bandwidth $W_0 = 9t$ in the absence of correlations [15]. **c**, The band parameters are experimentally accessible through the optical conductivity, where the peak position $\omega_{max}$ corresponds to $U$ and the half-width equals $W$. In the metallic case ($U = 0$) this transforms to a zero-frequency conductivity; for small, but finite scattering the width equals the scattering rate $\gamma$ of the Drude response while for bad metals the upper bound $W_0$ is approached. **d**, Schematic phase diagram of the Mott insulator with the sketched DOS for different regions. While in the high-temperature regime charge transport is dominated by incoherent thermal excitations and the Hubbard bands are strongly blurred, a well-defined charge gap opens in the Mott insulating state and a Drude peak appears on the metallic side at low temperatures. The boundary between the conducting regime and the fully gapped Mott state is the so-called quantum Widom line (dashed pink). The back-bending at low temperatures is nicely reproduced by DMFT calculations (green points). **e**, Zoom to the low-temperature region around the insulator-metal boundary. For $t = t'$ geometrical frustration suppresses magnetic order down to lowest temperatures; the large entropy of fluctuating spins causes a positive slope of the Mott MIT reminiscent of the Pomeranchuk effect in $^3$He. Up to the critical endpoint the phase transition is of first order while at higher temperatures the quantum Widom line defines a crossover. **f**, When geometrical frustration is reduced, antiferromagnetic order develops at low temperatures. The ordered state has less entropy than the metallic phase implying $dT_N/dp < 0$.

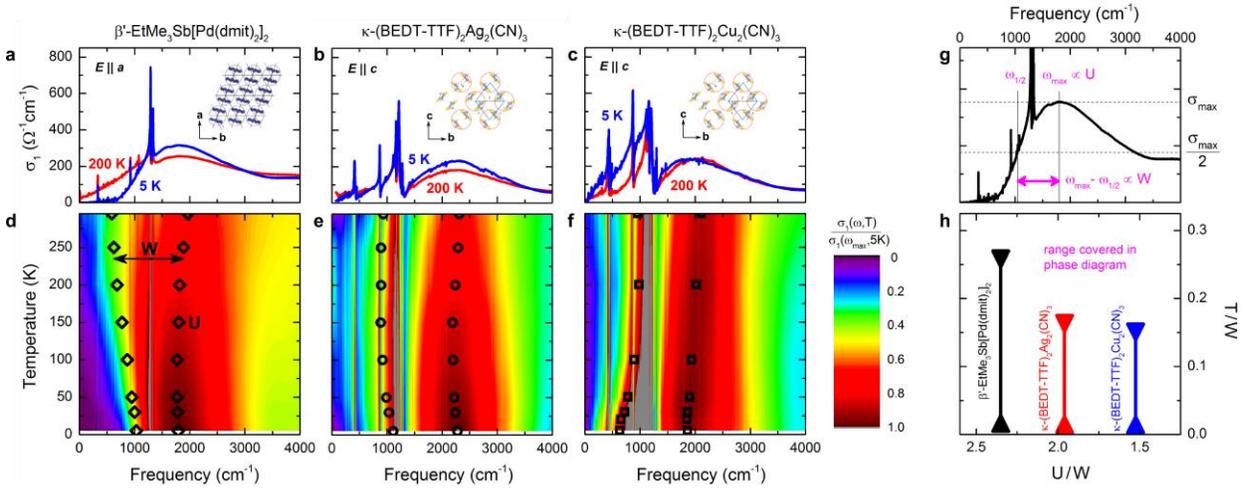

**Figure 2. Temperature evolution of the optical conductivity of three QSL compounds.**
**a-c**, The dominant feature that contains all information of the intrinsic Mott physics is the Mott-Hubbard band centered around 2000 cm$^{-1}$. At low frequencies narrow phonon modes appear on top. Note that the band shape and the low-frequency conductivity show distinct behavior for each compound which is related to the respective position in the phase diagram. **d-f**, The contour plots illustrate the temperature-dependent changes of the Mott-Hubbard band, where the open black symbols denote the maximum and half-maximum positions. **g,** The Coulomb repulsion $U$ corresponds to the band maximum position $\omega_{max}$ while the half-width at half maximum is proportional to the electronic bandwidth $W/\hbar = \omega_{max} - \omega_{1/2}$. **h**, The $U/W$ ratio at the lowest temperature defines the horizontal position in the phase diagram; the vertical bars correspond to the experimentally accessed temperature range (5 – 300 K) normalized to $k_B$ and the respective bandwidth. Hence, the $\beta'$-EtMe$_3$Sb[Pd(dmit)$_2$]$_2$ data extend over the largest $T/W$ range and correlations decrease via $\kappa$-(BEDT-TTF)$_2$Ag$_2$(CN)$_3$ and $\kappa$-(BEDT-TTF)$_2$Cu$_2$(CN)$_3$.

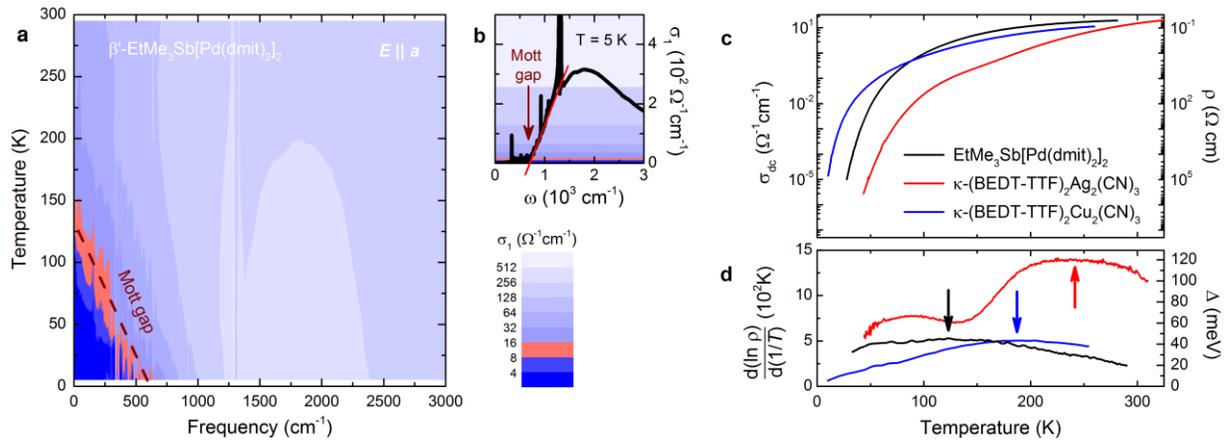

**Figure 3. Formation of the Mott gap in the optical and dc conductivity.**
**a**, The temperature- and frequency-dependent optical conductivity of $\beta'$-EtMe$_3$Sb[Pd(dmit)$_2$]$_2$ is displayed on a logarithmic blue-white scale. The region highlighted in red indicates the conductivity at the gap frequency, which is determined by linear extrapolation of the steepest slope (panel **b**). The Mott gap opens around 125 K and continuously grows upon cooling ($2\Delta = 650$ cm$^{-1}$ at T = 5 K). **c**, As these spin liquid compounds do not undergo any ordering, the dc conductivity drops with decreasing temperature without evidence of a sharp phase transition, in agreement with Refs. [26,27]. **d**, The temperature-dependent transport gap $\Delta(T)$ acquires a broad maximum corresponding to an inflection point in the Arrhenius plot of $\rho(T)$. For $\beta'$-EtMe$_3$Sb[Pd(dmit)$_2$]$_2$ (black) the transport gap peaks around 120 K coinciding with the opening of the Mott gap (panel **a**), thus it reflects the quantum Widom line. Also for $\kappa$-(BEDT-TTF)$_2$Cu$_2$(CN)$_3$ (blue) the maximum of $\Delta(T)$ agrees with the crossover temperature reported in Ref. [4].

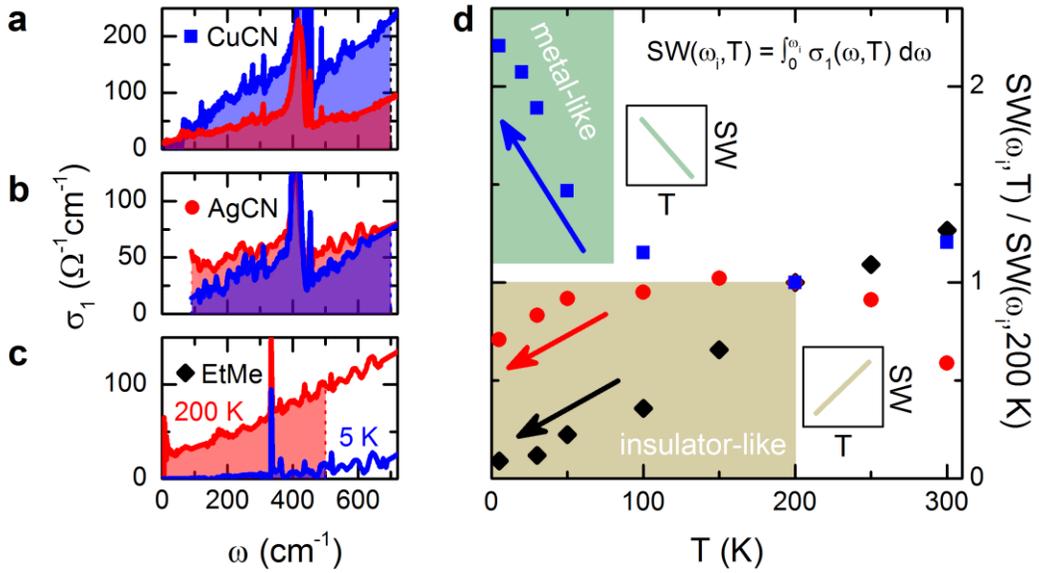

**Figure 4. Temperature dependence of the low-frequency spectral weight.**
**a-c**, The low-frequency optical conductivity exhibits distinct behaviour between 200 and 5 K. The shaded area indicates the corresponding spectral weight SW($\omega_i$,T) restricted to the region deep inside the Mott gap ($\omega_i$ = 500 cm$^{-1}$ for $\beta'$-EtMe$_3$Sb[Pd(dmit)$_2$]$_2$, 700 cm$^{-1}$ for $\kappa$-(BEDT-TTF)$_2$Ag$_2$(CN)$_3$ and $\kappa$-(BEDT-TTF)$_2$Cu$_2$(CN)$_3$). **d**, For $\beta'$-EtMe$_3$Sb[Pd(dmit)$_2$]$_2$ and $\kappa$-(BEDT-TTF)$_2$Ag$_2$(CN)$_3$ the SW freezes out, consistent with thermally-activated excitations across the gap diminishing at low temperatures. In contrast, for $\kappa$-(BEDT-TTF)$_2$Cu$_2$(CN)$_3$ the spectral weight strongly increases upon cooling, giving evidence for a novel absorption channel. The onset of this pronounced non-thermal enhancement appears at the same temperature as the back-bending of the quantum Widom line ($T_{back}$ = 70 K, cf. Fig. S1). The squares denote the ranges of metallic (increase of SW for T $\rightarrow$ 0) and insulating behaviour (decrease of SW for T $\rightarrow$ 0).

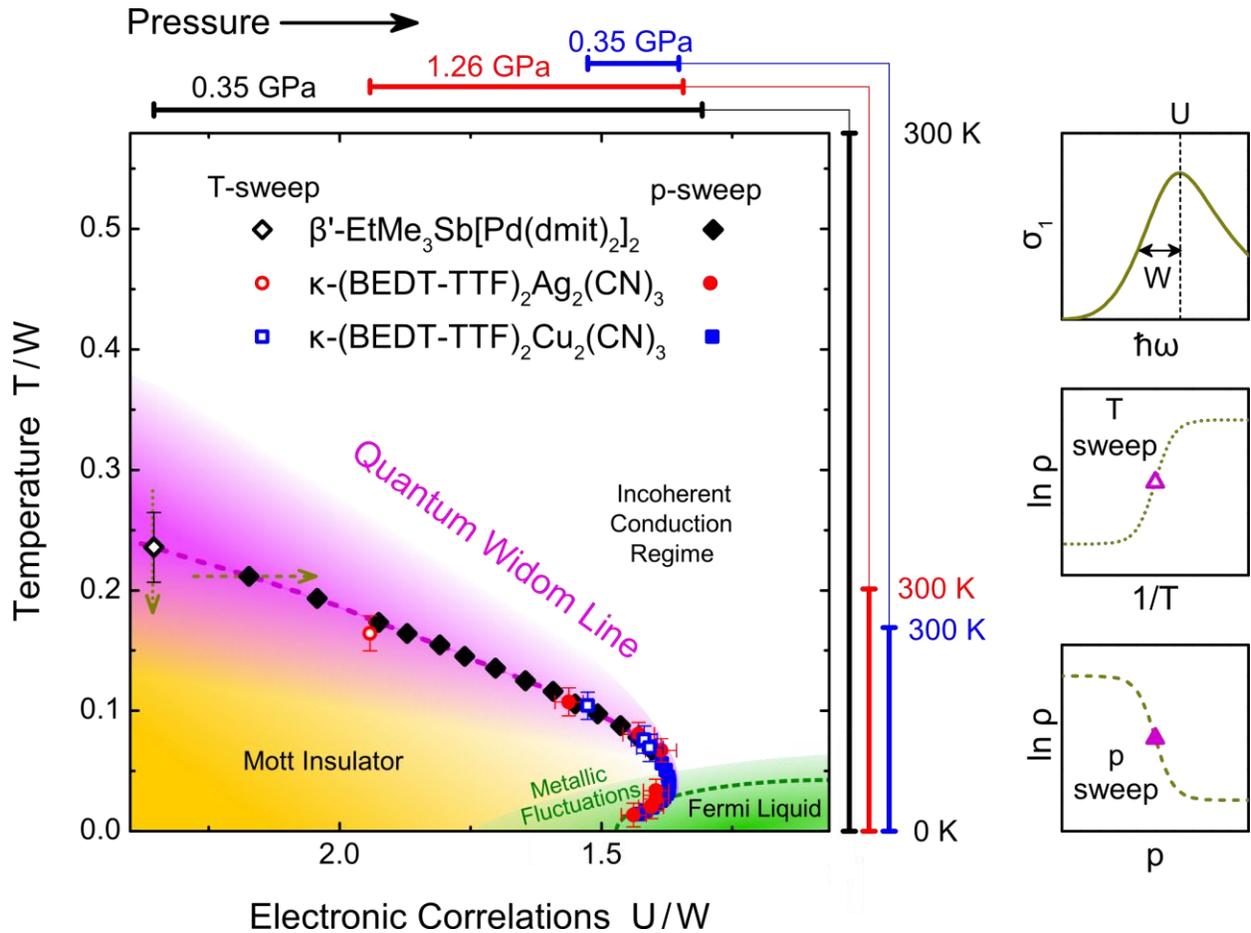

**Figure 5. Quantitative phase diagram of pristine Mott insulators.**
The temperature $T$ and Coulomb repulsion $U$ are normalized to the bandwidth $W$ extracted from optical spectroscopy; note that the direction of the bottom axis is reversed in order to mimic pressure. Since in these quantum spin liquids magnetic order is suppressed, the large residual entropy causes a pronounced back-bending of the quantum Widom line at low temperatures leading to metallic fluctuations (semi-transparent green area) in the Mott state close to the boundary. As the effective correlations decrease further, a metallic phase forms (green area) with Fermi liquid properties [34]. The universal phase diagram guided by the quantum Widom line is constructed on basis of our optical (Fig. 2) and transport (Figs. 3 and S9) data as well as the pressure-dependent transport studies [4,11] on $\beta'$-EtMe$_3$Sb[Pd(dmit)$_2$]$_2$ (black diamonds), $\kappa$-(BEDT-TTF)$_2$Ag$_2$(CN)$_3$ (red circles), and $\kappa$-(BEDT-TTF)$_2$Cu$_2$(CN)$_3$ (blue squares). Upon rescaling temperature (right scale) and pressure (top scale) the quantum Widom line is found to be universal for all compounds. The curvature, as well as the $T/W$ and $U/W$ values match well with theoretical calculations. On the right we illustrate how $U$ and $W$ are determined from our optical spectra and the quantum Widom line from the electrical resistivity $\rho(T,p)$ measured as a function of temperature (open symbols) and pressure (solid symbols), indicated by arrows in the main graph.

**Acknowledgments**
We thank K. Kanoda and R. Valenti for fruitful discussions.

Part of the work is supported by the Deutsche Forschungsgemeinschaft (DFG) via DR228/41-1, DR228/48-1. We also thank the Deutscher Akademischer Austauschdienst (DAAD) for support. This work was partially supported by JSPS KAKENHI Grant Number JP16H06346. We acknowledge the Russian Ministry of Education and Science (Program "5 top 100"). We also Acknowledge support from the Croatian Science Foundation project IP-2013-11-1011. Work in Florida was supported by the NSF Grant No. DMR-1410132.

**Contributions**
Most of the optical experiments and their analysis were conducted by A. P. with the help of M. B.. The THz measurements were performed by E. Z. and B. G.. Crystal growth and dc transport measurements on EtMe crystals were performed by R. K.. dc transport on AgCN and CuCN was measured by R. R. and A. L., respectively. The AgCN salts were grown by A. L., R. H., T. H., Y. Y. and G. S. and the CuCN crystals by A. L., R. H. and J. S.. Theoretical calculations were carried out by T.-H. L. and V. D. in communication with S. F.. The interpretation and draft of the manuscript were made by A. P. and M. D. who also conceived the project. All authors contributed to the discussion and the final manuscript.

**Competing financial interests:**
The authors declare no competing financial interests.

# Supplementary Information

# Quantum Spin Liquids Unveil the Genuine Mott State

A. Pustogow*, M. Bories, A. Löhle, R. Rösslhuber, E. Zhukova, B. Gorshunov, S. Tomić, J.A. Schlueter, R. Hübner, T. Hiramatsu, Y. Yoshida, G. Saito, R. Kato, T.-H. Lee, V. Dobrosavljević, S. Fratini and M. Dressel

**S1 Band Parameters: Definitions**

Strictly speaking, the Coulomb interaction $U$ referred to in the text and in Fig. 1 equates with the effective Coulomb interaction $U = U_0 - V_0$, where $U_0$ and $V_0$ are on-site and nearest-neighbor Coulomb repulsion, respectively. Upon moving an electron to its neighboring site a vacancy and a double occupancy are created, so-called holons and doublons. While the former reduces the nearest-neighbor repulsion, the latter enhances the on-site Coulomb energy, resulting in a total energy cost of $U = U_0 - V_0$. Empirically, the on-site term can be approximated with the strength of the intra-dimer transfer integrals $U_0 \approx 2t_d$, which is used in Table S1. There, we additionally compare the inter-dimer transfer integrals $t$, $t'$ as a measure of the geometrical frustration of the spin-liquid compounds under investigation. For completeness reasons we also list those of the related charge-transfer salts $\kappa$-(BEDT-TTF)$_2$Cu[N(CN)$_2$]Cl and $\kappa$-(BEDT-TTF)$_2$ Cu[N(CN)$_2$]Br; these are well-studied compounds with a significantly weaker frustration $t'/t$ that order antiferromagnetically or become metallic and even superconducting at low temperatures, depending on the $U/W$ ratio [S1,S2,S3,S4,S5]. The $U/D$ ratio obtained from *ab initio* density functional theory (DFT) and extended Hückel calculations based on the respective crystal structure indicates that correlations increase considerably when going from $\kappa$-(BEDT-TTF)$_2$Cu$_2$(CN)$_3$ (abbreviated CuCN) via $\kappa$-(BEDT-TTF)$_2$Ag$_2$(CN)$_3$ (abbreviated AgCN) to $\beta'$-EtMe$_3$Sb[Pd(dmit)$_2$]$_2$ (abbreviated EtMe) consistent with the experiments. Moreover, the bandwidth $W$ and Coulomb repulsion $U$ extracted from our optical experiments (Fig. 2) show the same trend. Deviations of the absolute value, however, stem from the limitations of the respective theoretical models.

| Compound | $t_d$ (meV) | $t$ (meV) | $t'$ (meV) | $t'/t$ | $U/t$ | $W_0 = 9t$ (meV) | $U/W_0$ | $(U/W)_{exp}$ |
|---|---|---|---|---|---|---|---|---|
| $\kappa$-(BEDT-TTF)$_2$Cu[N(CN)$_2$]Br | 200 | 78 | 33 | 0.42 | 5.1 | 500 | 0.8 | |
| $\kappa$-(BEDT-TTF)$_2$Cu[N(CN)$_2$]Cl | 200 | 73 | 32 | 0.44 | 5.5 | 476 | 0.84 | 1.5 |
| $\kappa$-(BEDT-TTF)$_2$Cu$_2$(CN)$_3$ | **200** | **55** | **45** | **0.83** | **7.3** | **450** | **0.89** | **1.52** |
| $\kappa$-(BEDT-TTF)$_2$Ag$_2$(CN)$_3$ | **264** | **53** | **48** | **0.90** | **10.5** | **454** | **1.16** | **1.96** |
| $\beta'$-EtMe$_3$Sb[Pd(dmit)$_2$]$_2$ | **454** | **28** | **26** | **0.92** | **32** | **244** | **3.72** | **2.35** |

**Table 1. Electronic parameters of several organic charge-transfer salts with different correlation strength.**

$\kappa$-(BEDT-TTF)$_2$Cu[N(CN)$_2$]Br is metallic and becomes superconducting below $T_c = 11.8$ K. $\kappa$-(BEDT-TTF)$_2$Cu[N(CN)$_2$]Cl is a Mott insulator that orders antiferromagnetically at $T_N = 25$ K; only 300 bar are sufficient to cross the insulator-to-metal transition. The spin liquid compounds $\kappa$-(BEDT-TTF)$_2$Cu$_2$(CN)$_3$ and $\kappa$-(BEDT-TTF)$_2$Ag$_2$(CN)$_3$ exhibit no magnetic order down to 20 mK, although the exchange coupling $J = 250$ K is rather strong. The coupling within the dimers $t_d$ is a measure of the Coulomb repulsion $U$. The overlap integrals $t$ and $t'$ characterize the frustration on the triangular lattice. The relative strength of correlations is given by $U/t$ and $U/W_0$. The last column gives parameters solely determined from our optical measurements. $U$ is experimentally determined from the mid-infrared maximum in the optical conductivity and the bandwidth $W$ is deduced from the width of the corresponding band (cf. Fig. 2). Because $W$ is defined from the full width at half maximum in of the density of states, it constitutes a lower bound to the actual bandwidth. Together with the sizable renormalization of the bandwidth expected due to electronic correlations [S6,S7,S8], this causes the extracted $U/W$ to be generally larger than the calculated $U/W_0$ [S3,S9,S10,S11,S12,S13,S14].

**S2 Construction of the Phase Diagram from Pressure-Dependent dc-Transport and Optics**

Making use of our conclusions from the optical spectroscopy results presented in Fig. 2 we relate the transport results of the three frustrated Mott insulators to each other and combine them to a universal quantum Widom line (QWL). In the *T-p* phase diagrams shown in Furukawa *et al.* [S15] the back-bending point of the QWL is located at a lower temperature for EtMe ($T_{back} < 35$ K) as compared to CuCN ($T_{back} = 70$ K) (Fig. S1). This agrees with the smaller bandwidth of EtMe providing evidence for the *T/W* scaling of the phase diagram. Considering that EtMe is located much deeper in the Mott insulating regime (cf. Fig. 2h) it is remarkable that the pressure at the back-bending point ($p_{back} \geq 3$ kbar) is similar to CuCN, consistent with the large compressibility of EtMe concluded from optical experiments. For AgCN, a similar compressibility but larger Coulomb repulsion compared to CuCN was concluded in a very recent study [S13], also consistent with our optical results. The overall agreement between optics, pressure-dependent transport and the theoretically determined band parameters strongly encourages our *T/W* vs. *U/W* scaling approach.

Fig. S1 illustrates the scaling procedure to connect the QWL of EtMe and CuCN based on the pressure-dependent dc transport results by Furukawa *et al.* [S15] and the data shown in Figs. 3d and S9. Due to different bandwidth and electronic correlations each compound covers a distinct range in the phase diagram. Upon rescaling the temperature and pressure axes by *U* and *W*, the intrinsic Mott physics is revealed and all data points collapse on a universal quantum Widom line. A good match between the EtMe and CuCN data sets can be reached for $2 < T_{EtMe}/T_{CuCN} < 4.5$. The best fit, however, is achieved for a bandwidth ratio of 3.4, when also the normalized resistance values, i.e. the color scale in Fig. 1d and f from Ref. [S15], are in agreement. For AgCN, the QWL was extracted from the pressure-dependent resistivity reported in Ref. [S13] and the ambient pressure result from Fig. 3d, and then similarly scaled to match with the EtMe and CuCN data.

Taking CuCN as our reference material, we rescale the temperature and pressure of EtMe by factors 3.4 and 6, respectively; AgCN has a temperature scaling factor of 1.2 compared to CuCN, almost equal compressibility and is shifted to the left on the horizontal axis due to the larger *U*, in good agreement with Ref. [S13]. The ambient conditions are located at -1.66 GPa for EtMe and -0.83 GPa for AgCN on the CuCN pressure scale implying the following linear scaling relations:

$$p_{CuCN} = -0.83 GPa + 0.95 p_{AgCN} = -1.66 GPa + 6 p_{EtMe} . \qquad (S1)$$

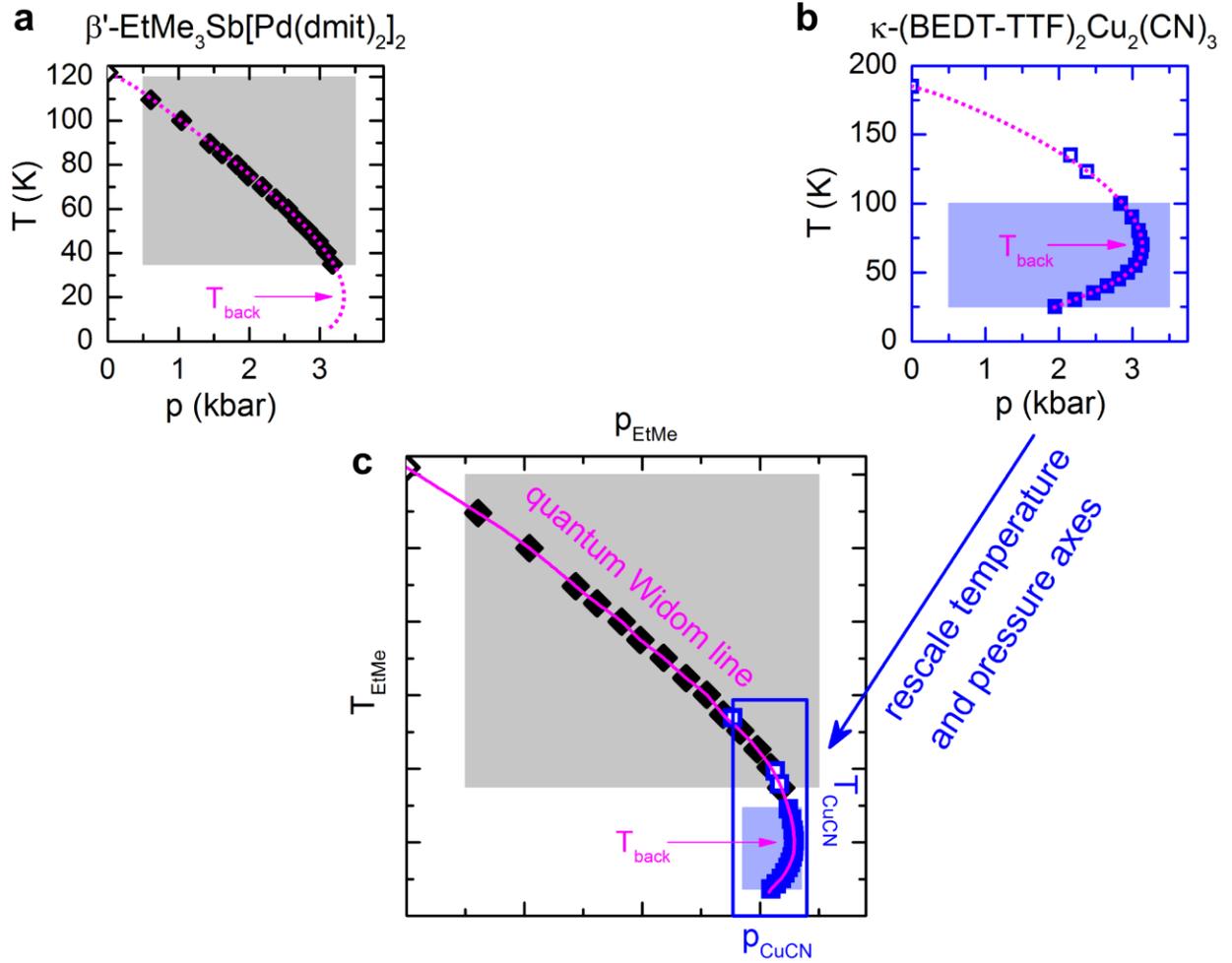

**Figure S1. Rescaling procedure to connect the quantum Widom lines of $\beta'$-EtMe$_3$Sb [Pd(dmit)$_2$]$_2$ (black) and $\kappa$-(BEDT-TTF)$_2$ Cu$_2$(CN)$_3$ (blue).**

**a,b**, The temperature and pressure range covered by the dc transport experiments of Furukawa *et al.* [S15] is highlighted in grey and blue, respectively; the solid symbols correspond to the quantum Widom line (QWL) determined by pressure sweeps at constant temperature. The dotted magenta lines illustrate the interpolation of the QWL. The QWL coincides with the maximum of the transport gap $\Delta(T)$ (open symbols, cf. Fig. 3d). **c**, Using the back-bending at $T_{back}$ as a fixed point, the two data sets can be connected smoothly by stretching and squeezing of the respective axes, resulting in a universal QWL (solid magenta line). As indicated by the blue rectangle, temperature and pressure of $\kappa$-(BEDT-TTF)$_2$Cu$_2$(CN)$_3$ are scaled down by factors of 3.4 and 6 compared to $\beta'$-EtMe$_3$Sb[Pd(dmit)$_2$]$_2$, respectively. This implies that $\beta'$-EtMe$_3$Sb[Pd(dmit)$_2$]$_2$ has a smaller bandwidth, larger compressibility and its ambient pressure position is located more *left* in the phase diagram than for $\kappa$-(BEDT-TTF)$_2$Cu$_2$(CN)$_3$, consistent with the optical results from Fig. 2.

Page 22 of 40

Due to the different experimental approach of Shimizu *et al.* [S13], who measured temperature sweeps at constant pressure rather than pressure sweeps at constant temperature, the QWL determination was not as accurate for AgCN, as indicated by the error bars in Fig. 5. Note that they concluded an absolute pressure shift of 0.6 GPa between AgCN and CuCN while our procedure results in a shift of approximately 0.9 GPa, as shown in equation (S1). Their assumption, however, is based on pressure-dependent transport studies of CuCN employing oil as pressure medium [S17]. It is known that such setups suffer from a pressure loss at low temperatures when the oil freezes, making the pressure determination at low temperatures less accurate. On the other hand, the data of Furukawa *et al.* [S15] on CuCN were measured in a He gas cylinder where the applied pressure can be adjusted and determined very accurately even at low temperatures. This explains the different pressure scale compared to Kurosaki *et al.* [S17], where superconductivity sets in at 0.35 GPa while the insulator-superconductor phase boundary of CuCN is located at 0.125 GPa in Ref. [S15].

The unified phase diagram presented in Fig. 5 based on the bandwidth $W$ and Coulomb repulsion $U$ determined by optical spectroscopy (cf. Fig. 2, Fig. S5) shows excellent agreement between the optical and transport investigations. Comparing the experimental QWL (Fig. 5) with the dynamical mean field theory (DMFT) calculations employing the iterated perturbation theory solver (Fig. 1d) yields similar $U/W_0$ values at the reversal point. Therefore, the critical correlations needed to drive the metal into a Mott insulator are reproduced very well. Also the vertical position of the QWL, i.e. the optically determined $T/W$ values of AgCN and CuCN at ambient pressure (245 K $k_B$ / $W_{AgCN}$ = 0.14 and 185 K $k_B$ / $W_{CuCN}$ = 0.11, cf. Fig. 3 **E**), is close to the $T/W_0$ scale of the theoretical data (($k_B T_{QWL}/W$)$_{AgCN}$ = 0.24 and ($k_B T_{QWL}/W$)$_{CuCN}$ = 0.14). Keeping in mind the general overestimation of the critical $T/W_0$ and $U/W_0$ related to the iterated perturbation theory, these values are in good consensus. Thus, in addition to the excellent agreement of the qualitative shape of the QWL, there is also quantitative accordance between the optical, transport and theoretical results. While there is quantitative consistency for AgCN and CuCN, the $T/W$ range covered by EtMe is larger than inferred from the optically determined bandwidth. Note, however, that the rescaling of transport data involves considerable insecurity since the QWL reversal of EtMe is not captured by the data of Ref. [S15]. Having a complete data set would certainly improve the quantitative accordance for this compound. Nevertheless it is remarkable that, for AgCN and CuCN, both the $T/W$ and $U/W$ scales agree well with the scaled transport results and DMFT calculations, and also the ambient pressure position $(U/W)_{EtMe}$ shows quantitative agreement on the common $U/W$ scale.

## S3 Optical Conductivity

Figs. S2 and S3 illustrate the determination of the optical conductivity from the reflectivity $R$ along both in-plane crystal axes. The phase shift $\Phi$ of the complex reflectivity $r = \sqrt{R}e^{i\Phi}$ is computed from the broad-band $R(\omega)$ data using the Kramers-Kronig relations, which allows for direct calculation of the complex refractive index and conductivity [S16].

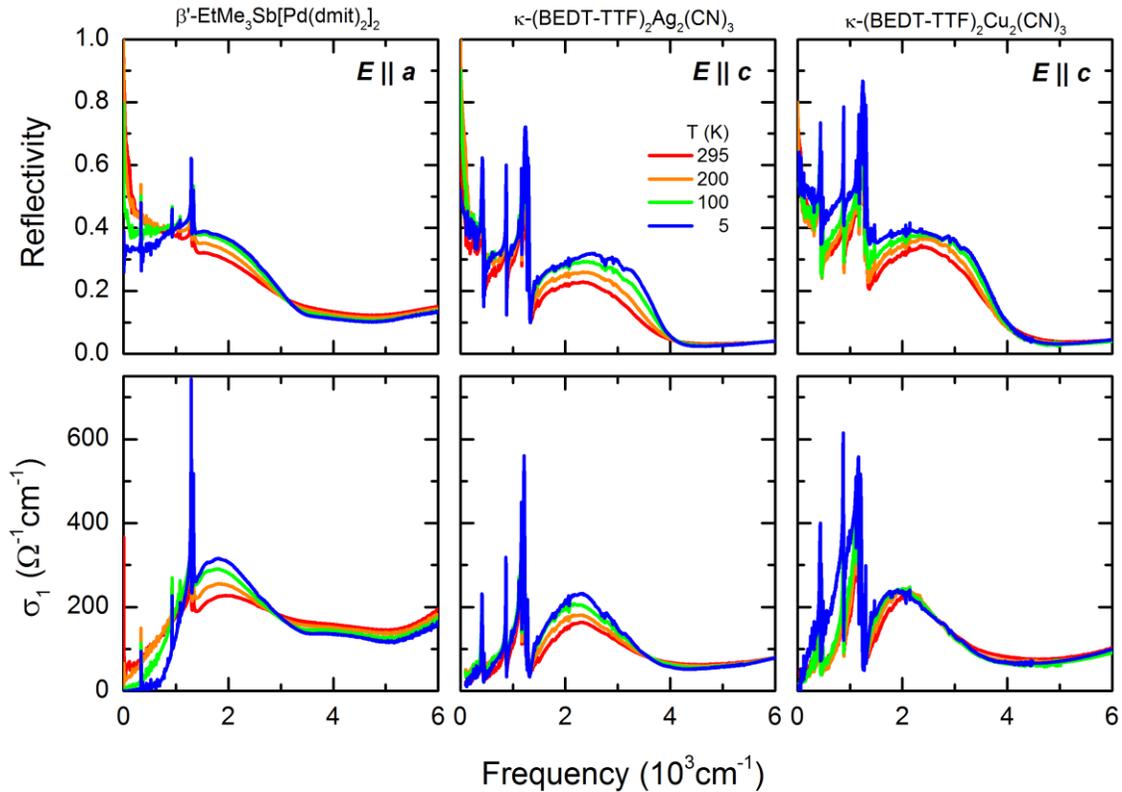

**Figure S2. Optical reflectivity (upper panels) and conductivity (lower panels) along the in-plane direction with largest dc conductivity.**

The optical conductivity of $\beta'$-EtMe$_3$Sb[Pd(dmit)$_2$]$_2$ ($E \parallel a$), $\kappa$-(BEDT-TTF)$_2$Ag$_2$(CN)$_3$ ($E \parallel c$) and $\kappa$-(BEDT-TTF)$_2$Cu$_2$(CN)$_3$ ($E \parallel c$) was calculated from the broad-band reflectivity data using the Kramers-Kronig relations. For convenience, only selected temperatures are plotted.

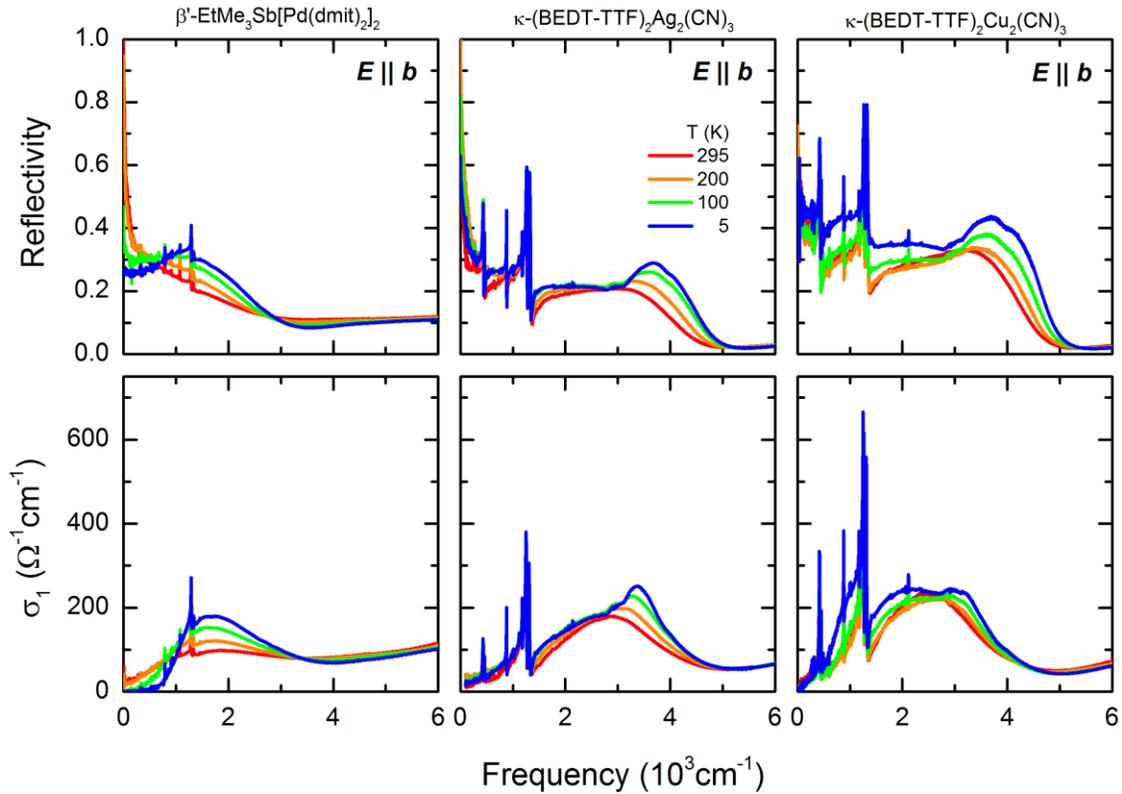

**Figure S3. Optical reflectivity (upper panels) and conductivity (lower panels) along the in-plane direction with lowest dc conductivity.**

The side band occurring for $E \parallel b$ in $\kappa$-(BEDT-TTF)$_2$Ag$_2$(CN)$_3$ and $\kappa$-(BEDT-TTF)$_2$Cu$_2$(CN)$_3$ results from intra-dimer charge transfer. Apart from the absolute values, phonons and molecular vibrations, the low-frequency optical conductivity shows qualitatively similar properties and temperature dependencies as for $E \parallel c$ (Fig. S2). Also for $\beta'$-EtMe$_3$Sb[Pd(dmit)$_2$]$_2$ the physical properties are the same for $E \parallel a$ and $E \parallel b$.

In Fig. S4 we plot the temperature-dependent optical conductivity $\sigma_1(\omega)$ for both in-plane polarizations of the three compounds under inspection, EtMe, AgCN and CuCN. In their overall appearance, the spectra are very similar for both polarizations with the charge-transfer band around 2000 cm$^{-1}$. In the two $\kappa$-salts, however, the peak splits into a low- and high-energy part for the polarization $E \parallel b$. The latter corresponds to intra-dimer excitations while the former one is related to the charge transport between the dimers; due to strong Coulomb repulsion the conduction band is split into the lower and upper Hubbard bands with transitions between them right around 2000 cm$^{-1}$ [S1,S5,S18]. Due to the overlapping features, the determination of $\omega_{max}$ of the inter-dimer part was less accurate as for the $c$-direction where the intra-dimer charge transfer is less pronounced. In the case of the EtMe compound the intra-dimer excitation band appears at much higher frequencies (around 4000 cm$^{-1}$ for $E \parallel a$) because of its stronger dimerization.

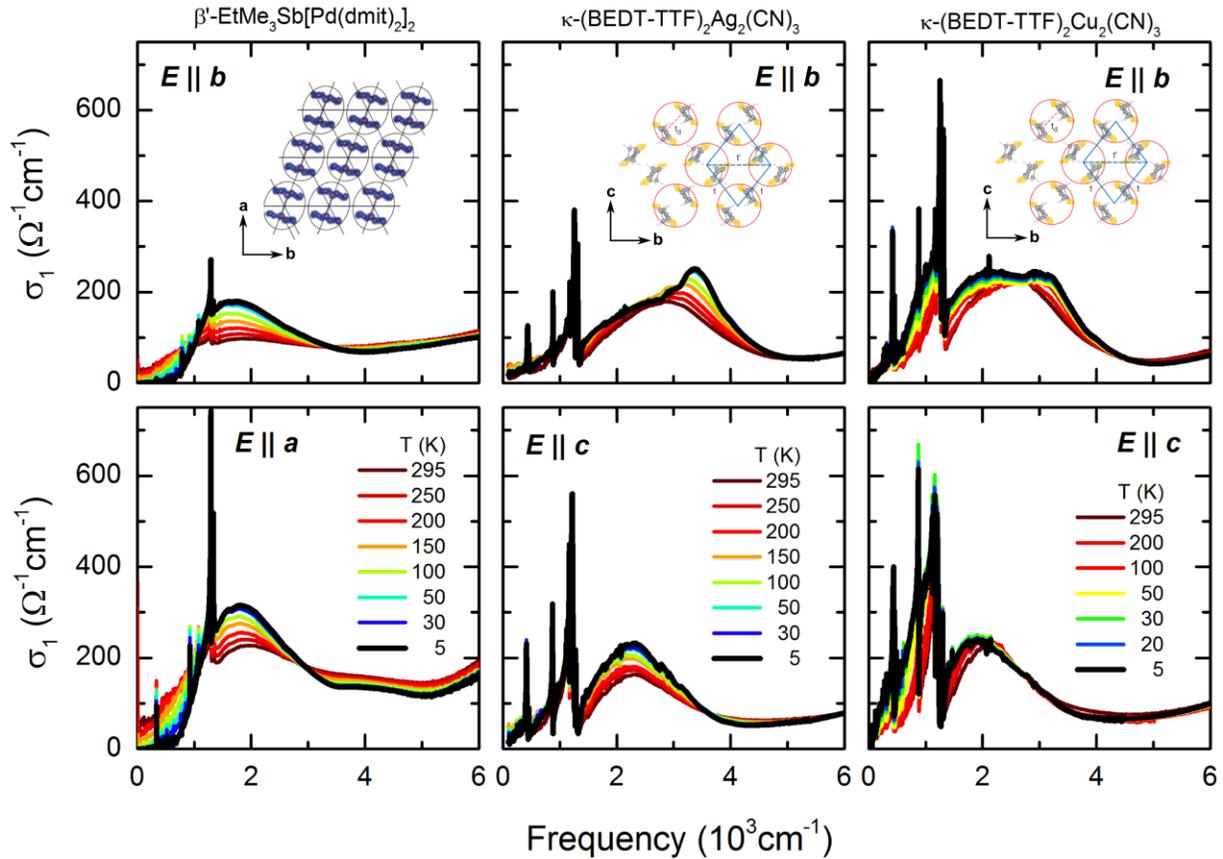

**Figure S4. Optical conductivity of the studied compounds along both in-plane polarizations.**

All spectra are dominated by the electronic transition from the lower to the upper Hubbard bands. The side band occurring for $E \parallel b$ in $\kappa$-(BEDT-TTF)$_2$Cu$_2$(CN)$_3$ and $\kappa$-(BEDT-TTF)$_2$Ag$_2$(CN)$_3$, and at 4000 cm$^{-1}$ for $E \parallel a$ in $\beta'$-EtMe$_3$Sb[Pd(dmit)$_2$]$_2$, results from intra-dimer charge transfer. Apart from phonons and molecular vibrations, the low-frequency optical conductivity shows qualitatively similar properties and temperature dependencies for both polarizations.

We also extracted the optical parameters for $E \parallel b$ which is plotted in Fig. S5 (empty symbols) together with the data from Fig. 2d-f (filled symbols). Most importantly, the $U/W_0$ ratio in panel c shows qualitatively similar behavior for both crystal directions. At low temperatures the thermal broadening of the bandwidth of EtMe and AgCN nicely follows the theoretically expected temperature dependence [S8] $\sqrt{W_0^2 + aT^2}$. Additional smearing at elevated temperatures might be related to thermal population of low-frequency inter-molecular vibrations.

At the lowest temperature, the absolute *U/W* value is slightly larger along E ∥ b implying stronger correlations and, thus, a position more *left* in the phase diagram. However, this anisotropy is substantially smaller than the insecurity in combining the Widom lines from transport measurements (see Section S2). The $(U/W)_{exp}$ values in Table 1 are taken from E ∥ c due to the generally smaller error bar.

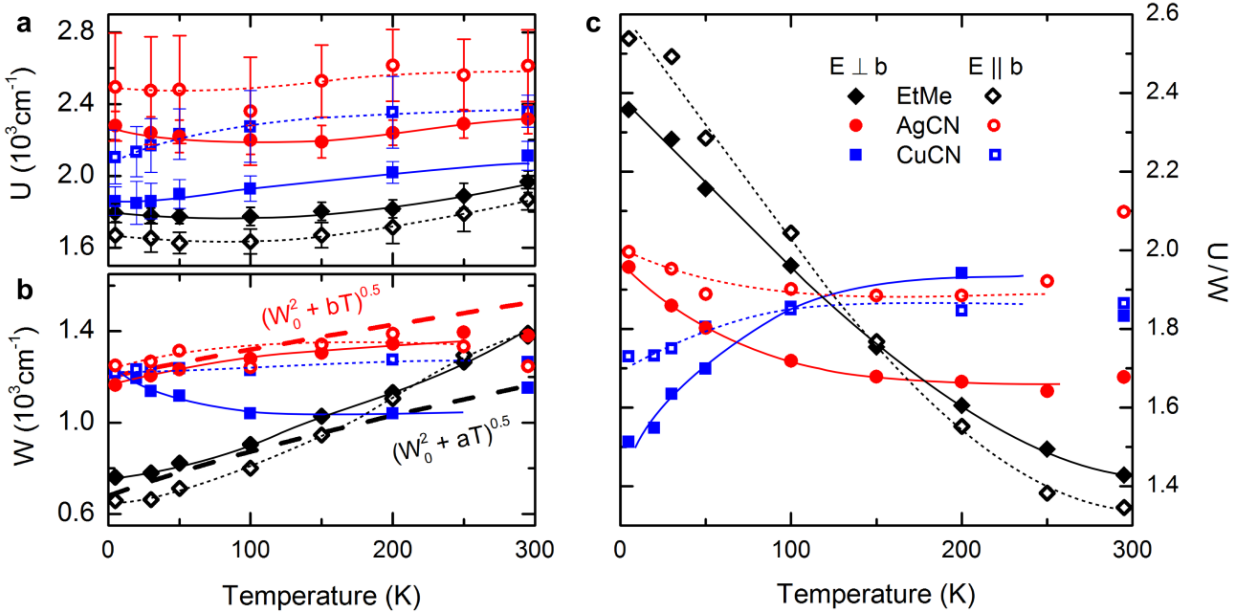

**Figure S5. The optical parameters for both in-plane directions *E* ∥ *b* (open symbols) and *E* ⊥ *b* (solid symbols).**

**a**, Temperature dependence of the maximum of the Mott-Hubbard band, which is proportional to *U*. **b**, Half-width at half-maximum of the Mott-Hubbard band $\omega_{max} - \omega_{1/2}$ proportional to *W*. The thick dashed lines correspond to a low-temperature approximation of the bandwidth broadening expected for pure thermal excitations [S8], which works reasonably well for $\beta'$-EtMe$_3$Sb[Pd(dmit)$_2$]$_2$ and $\kappa$-(BEDT-TTF)$_2$Ag$_2$(CN)$_3$ up to $T \approx 150$ K. **c**, Ratio of these quantities *U* and *W* as a function of temperature [cf. Fig. 2d-f]. For both crystallographic directions we find a comparable *T* dependence. Due to overlapping inter- and intra-band transitions, the error bar is larger for *E* ∥ *b*, as indicated in panel **a**. Dashed and solid lines are guides to the eye.

## S4 Low-Energy Electrodynamic Response

Metals and insulators are commonly defined according to the temperature-dependence of their resistivity. In a metal, the main changes of resistivity are related to the scattering rate $\gamma$, i.e. $\rho \propto \gamma \propto \sigma_{dc}^{-1}$, which increases with temperature and thus enhances (reduces) the resistivity (dc conductivity). For an insulator, the dominant contribution to charge transport is thermal activation; unlike in a metal the conductivity increases with temperature. As sketched in Fig. S6, a similar behavior is expected for the low-frequency optical conductivity $\sigma_1(\omega, T)$ at energies well below the gap and scattering rate for insulators and metals, respectively.

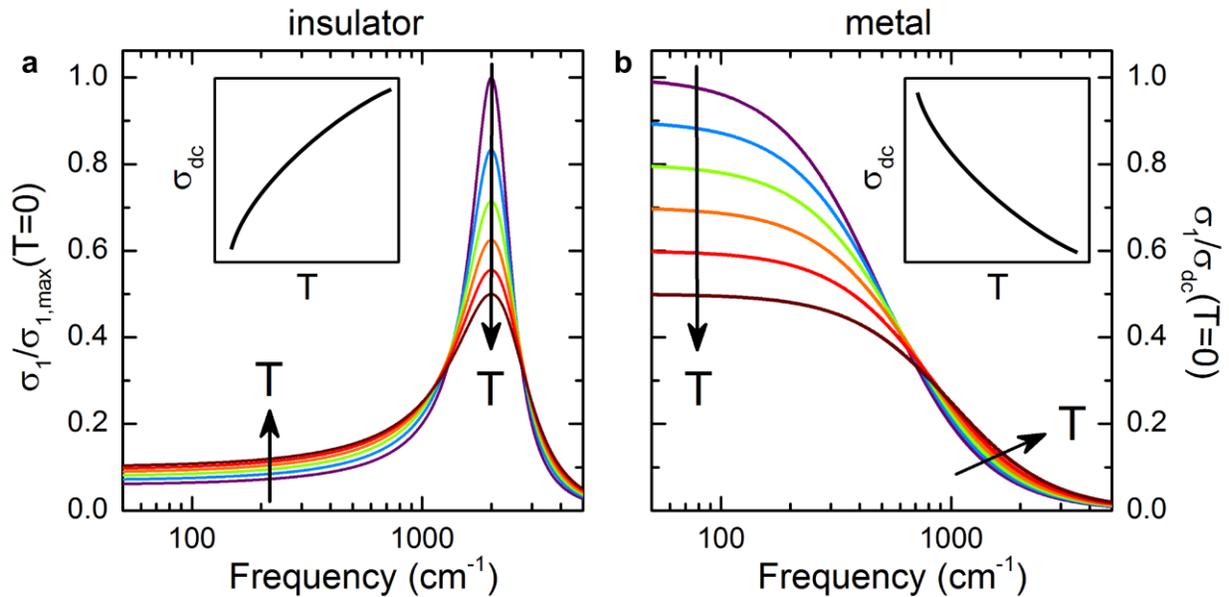

**Figure S6. Characteristic spectroscopic and dc transport properties of insulators and metals.**

**a,** For an insulator, the band is located at finite frequency and the conductivity within the gap is very small. Upon increasing the temperature the spectral weight is shifted from the maximum to the (high- and) low-energy region causing an enhancement of the low-frequency conductivity. **b,** Since for a metal the Drude response is centered at $\omega = 0$, the low-energy contribution is suppressed with temperature. In the limit of very small frequencies the optical conductivity approaches the dc conductivity (insets), i.e. $\sigma_1(\omega=0) = \sigma_{dc} = \rho^{-1}$. Thus, the low-energy spectral weight SW(T) usually reveals a similar temperature evolution as $\sigma_{dc}(T)$; this is also illustrated in Fig. 4d. For an insulator the conductivity is thermally activated and thus increases with temperature; the opposite behavior is found for a metal, where scattering impedes charge transport.

As intensely discussed in the main part, for EtMe and AgCN insulating behavior is consistently found in the dc and low-frequency optical conductivity. For CuCN, however, the sub-gap spectral weight exhibits an anomalous increase upon cooling reminiscent of metallic behavior despite insulating transport properties. We interpreted this anomaly as metallic fluctuations appearing in the vicinity of the insulator-metal phase boundary.

Although there is some anisotropy in the dc response and optical conductivity, the overall temperature dependence of charge transport is similar for both in-plane crystal axes. As shown in Fig. S7a, in CuCN and AgCN the low-frequency optical conductivity is higher along the $c$-direction compared to $E \parallel b$, similar to the dc results [S19,S20]. This anisotropy is also expressed in the maximum position of the band; similar to the transport gap, $U$ is larger for $E \parallel b$ (Fig. S5a). This observation is explained by the fact that the bandwidth is largest along the $c$-direction. There is also small anisotropy for EtMe. The overall temperature evolution of the low-frequency response (Fig. S7b,c), however, is robust and does not depend on the polarization which assures our assumption that we probe the intrinsic behavior of the two-dimensional electron gas subject to strong correlations. Having two independent measurements strengthens our conclusion.

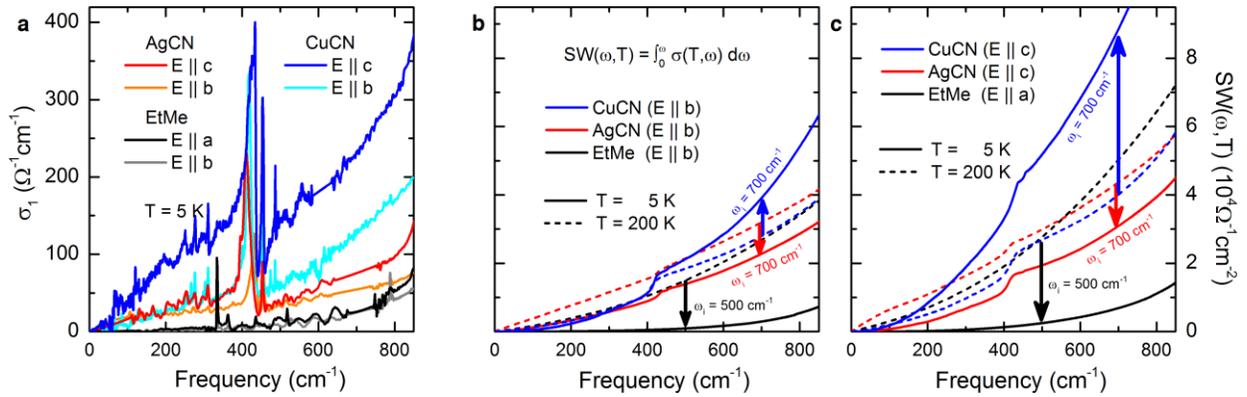

**Figure S7.** Polarization and temperature dependence of the low-frequency absorption. **a**, Low-frequency conductivity of $\beta'$-EtMe$_3$Sb[Pd(dmit)$_2$]$_2$, $\kappa$-(BEDT-TTF)$_2$ Ag$_2$(CN)$_3$ and $\kappa$-(BEDT-TTF)$_2$Cu$_2$(CN)$_3$ for both polarizations shown at the lowest measured temperature. **b,c**, Although the optical conductivity is anisotropic, there is no significant difference in the temperature dependence of the spectral weight, as indicated by the arrows. While it decreases for the former two compounds, it is enhanced for the latter one upon cooling, which is unexpected for an insulator (cf. Fig. S6). The arrows denote the frequency $\omega_i$ at which the temperature dependence of the spectral weight was compared in Figs. 4 and S11a.

## S5 Theoretical Calculations

In Fig. S8 we present the DOS (panel a) and optical conductivity (panel b) calculated by the continuous time quantum Monte Carlo (CTQMC) method. As expected, and in agreement with the experimental spectra presented in Fig. 2 and S4, the maximum of $\sigma_1(\omega)$ shifts to higher energies as $U$ increases. Importantly, the asymmetric shape of the Mott-Hubbard band observed in experiment is nicely reproduced. As shown in the inset of b, the maximum position of the optical conductivity is related linearly to the Coulomb repulsion $U$ establishing our approach to extract the electronic interactions from the experimentally observed Mott-Hubbard band. The theoretical $\sigma_1$ scale is calculated in units of the Ioffe-Regel-Mott (IRM) limit $\sigma_{1,IRM} = e^2/(hd)$, where h is Planck's constant and $d$ the inter-layer distance, which allows for quantitative comparison with our experimental data. Here, the IRM limit is around 210 $\Omega^{-1}$cm$^{-1}$ for EtMe and 260 $\Omega^{-1}$cm$^{-1}$ for the two $\kappa$-compounds [S10,S12,S21]. Thus, the experimental values are reproduced by our CTQMC calculations up to the same order of magnitude.

Note, that we used a Bethe lattice here while a triangular lattice was assumed for the determination of the quantum Widom line shown in Fig. 1d. As a result, the absolute values of $U/W_0$ might have an offset and different critical ratio at the Mott MIT; the general Mott physics, however, is captured by both approaches, as demonstrated in Ref. [S22].

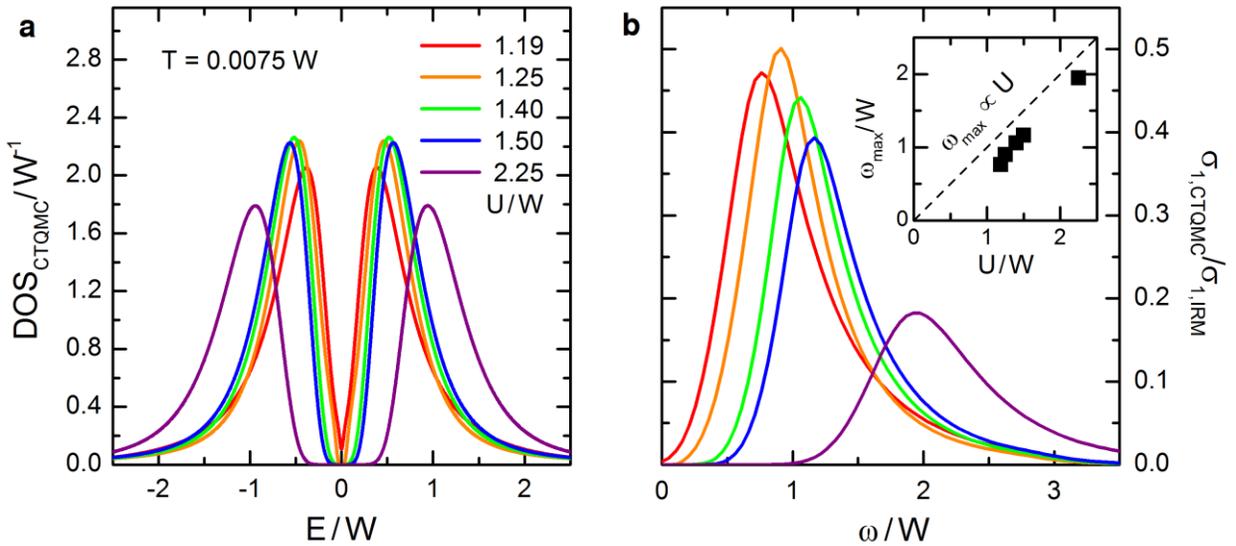

**Figure S8. Density of states and optical conductivity calculated by continuous time quantum Monte Carlo methods.**
The asymmetric shape of the Mott-Hubbard bands observed in our optical experiments (Fig. 2) is reproduced both in the DOS (**a**) and conductivity $\sigma_1$ (**b**) for different correlation strength $U/W$ as indicated. The calculated optical conductivity is plotted in units of the Ioffe-Regel-Mott limit.

The assumed $T/W = 0.0075$ corresponds to the low-temperature limit of our experiments. *Inset:* The linear relation between the maxima of optical conductivity and the Coulomb repulsion $U$ reassures our analysis based on Fig. 2g. Note, that due to the implicit assumptions of the theoretical model, the band edge is generally much steeper than in the materials under study. Here we utilized a Bethe lattice, while for the iterated perturbation theory applied in Fig. 1d a triangular lattice was assumed, leading to different $U/W$ values.

**S6 Mott Physics Above the Antiferromagnetic State in $\kappa$-(BEDT-TTF)$_2$Cu[N(CN)$_2$]Br$_x$Cl$_{1-x}$**

In the following, the quantum spin liquids are compared to the less-frustrated $\kappa$-(BEDT-TTF)$_2$ Cu[N(CN)$_2$]Br$_x$Cl$_{1-x}$ series (abbreviated $\kappa$-Br$_x$Cl$_{1-x}$) that shows magnetic order in the insulating state at low temperatures. While $T_N \approx 25$ K for $x = 0$ and $x = 0.4$, the compound with $x = 0.7$ is subject to an insulator-metal transition at $T_{MI} \approx 50$ K and becomes superconducting below $T_c \approx 12$ K [S1,S3,S4].

In Fig. S9a we plot the phase diagram of the three $\kappa$-Br$_x$Cl$_{1-x}$ compounds based on their transport [S15,S4] and optical properties [S1,S3]. Due to similar bandwidth we use the same temperature and pressure scale as for CuCN and indicate the respective positions for each compound. At elevated temperatures Mott physics is dominant and the $\kappa$-Br$_x$Cl$_{1-x}$ compounds follow the same QWL (black line) as CuCN. At low temperatures, however, antiferromagnetic (AFM) order forms, causing a different behavior than in the QSL compound. In particular, the Pomeranchuk-like back-bending of the QWL (open white squares indicate the phase boundary of CuCN) is suppressed by antiferromagnetic fluctuations such that the insulator-metal boundary (dashed pink) acquires a negative slope close to the AFM state. Note that in the presence of strong geometrical frustration in CuCN the metallic state is stabilized with respect to the Mott state up to larger correlations $U/W$, i.e. smaller pressure, than in the less-frustrated $\kappa$-Br$_x$Cl$_{1-x}$ series. Thus, AFM and its fluctuations suppress the Fermi liquid and affect thermodynamics more strongly than the pure interplay of charge excitations and Coulomb repulsion.

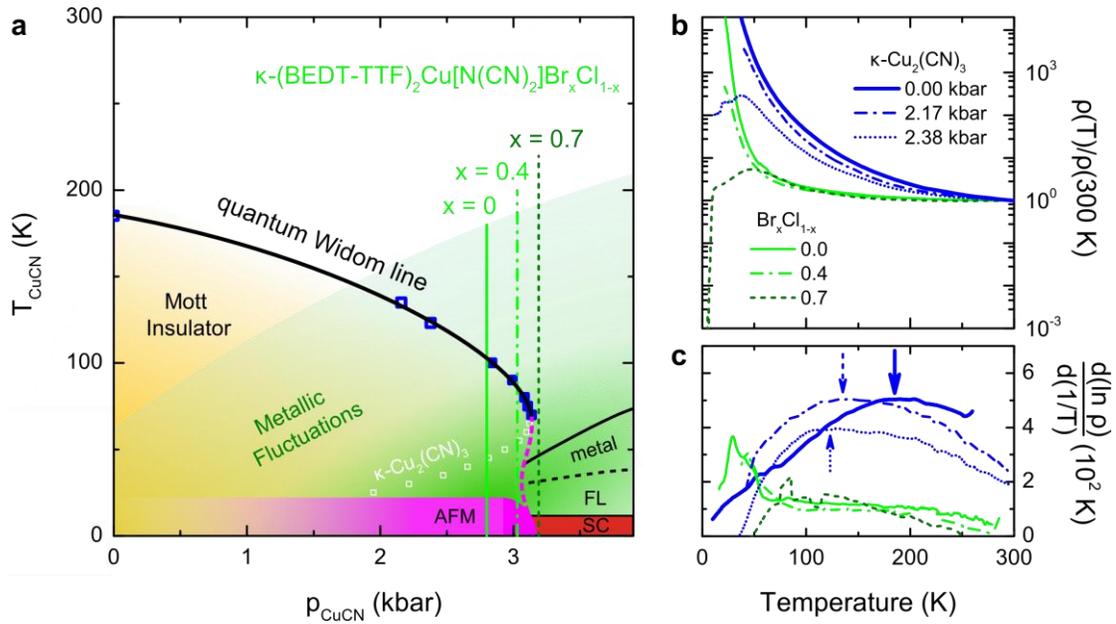

**Figure S9. Phase diagram the antiferromagnetic $\kappa$-(BEDT-TTF)$_2$Cu[N(CN)$_2$]Br$_x$Cl$_{1-x}$.**
**a**, The black line corresponds to the common high-temperature quantum Widom line of $\kappa$-(BEDT-TTF)$_2$Cu$_2$(CN)$_3$ (blue squares, cf. Figs. 5, S1) and $\kappa$-(BEDT-TTF)$_2$Cu[N(CN)$_2$]Br$_x$Cl$_{1-x}$. The solid, dashed-dotted and dashed green lines refer to the ambient pressure positions of the compounds with $x$ = 0, 0.4 and 0.7, respectively [S1,S3,S4,S15]. As indicated in semi-transparent green color, metallic fluctuations appear both in the Mott state and in the incoherent regime above and right of the quantum Widom line (see discussion of Fig. S11a). Since geometrical frustration is much smaller than in $\kappa$-(BEDT-TTF)$_2$Cu$_2$(CN)$_3$, antiferromagnetic (AFM) order sets in below $T_N \approx$ 25 K for $x \leq$ 0.4 as indicated in pink. The associated antiferromagnetic fluctuations suppress the back-bending of the quantum Widom line (QWL) and cause a negative slope of the insulator-metal boundary in the vicinity of the ordered state (dashed pink line). The open white squares indicate how the low-temperature phase boundary would form in absence of AFM, realized by the geometrically frustrated $\kappa$-(BEDT-TTF)$_2$Cu$_2$(CN)$_3$. Thus, magnetic order suppresses the metallic phase up to higher pressure (i.e. smaller $U/W$ ratio) compared to a pure Mott insulator. **b**, The resistivity data reveal that increasing the Br-content effectively enhances the bandwidth leading to metallic and superconducting behavior at low temperatures. We also plot the pressure-dependent resistivity of CuCN (blue) for p = 0, 2.17 and 2.38 kbar. **c**, The logarithmic resistivity derivative of the $\kappa$-(BEDT-TTF)$_2$Cu[N(CN)$_2$]Br$_x$Cl$_{1-x}$ compounds is qualitatively different from the quantum spin liquids. A pronounced peak occurs around $T_N$. The transport gap maximum of CuCN (open blue squares in **a**) shifts to lower temperature upon pressure.

The transport properties shown in Fig. S9b reveal the typically smaller resistivity of κ-Br$_x$Cl$_{1-x}$ in accordance with the location much closer to the metallic state due to smaller correlations $U/W$. This effect is seen very clearly in frame (c) as the transport gap is significantly smaller than in CuCN. In the vicinity of its reversal point the QWL acquires a nearly vertical slope over a broad temperature range (for $x = 0$, see Fig. 1e from Ref. [S15]). Hence, the contribution to the transport gap related with Mott physics manifests as a broad plateau above 50 K. The bandwidth tuning effect becomes obvious as (i) the size of the transport gap is much smaller than in the three quantum spin liquid compounds under study with stronger correlations, and (ii) the high-temperature tail of the plateau is suppressed for larger Br-content $x$. When entering the AFM state, however, the resistance shoots up drastically which expresses as a sharp maximum centered at $T_N$ in the logarithmic derivative. The onset happens well above the magnetic transition, indicating magnetic fluctuations for $T_N < T < 50$K.

To compare the electrodynamic properties in the context of Mott physics, we confine ourselves to a temperature right above the onset of magnetic fluctuations. In Fig. S10a we plot the optical conductivity of the three quantum spin liquids and the three κ-Br$_x$Cl$_{1-x}$ compounds at $T = 50$ K $> T_N$. To facilitate comparison, we renormalized the Mott-Hubbard band to the maximum conductivity $\sigma_{1,max}$ and frequency $\omega_{max}$, respectively. Additionally, we smoothed the data as the frequency shift moves the vibrational features differently for each compound making the picture confusing. As a general trend, the in-gap conductivity is enhanced for smaller correlations. The optical conductivity calculated by the continuous time quantum Monte Carlo method shown in Fig. S8b was normalized the same way and, similarly, exhibits a pronounced broadening of the band for reduced $U/W$ (Fig. S10b).

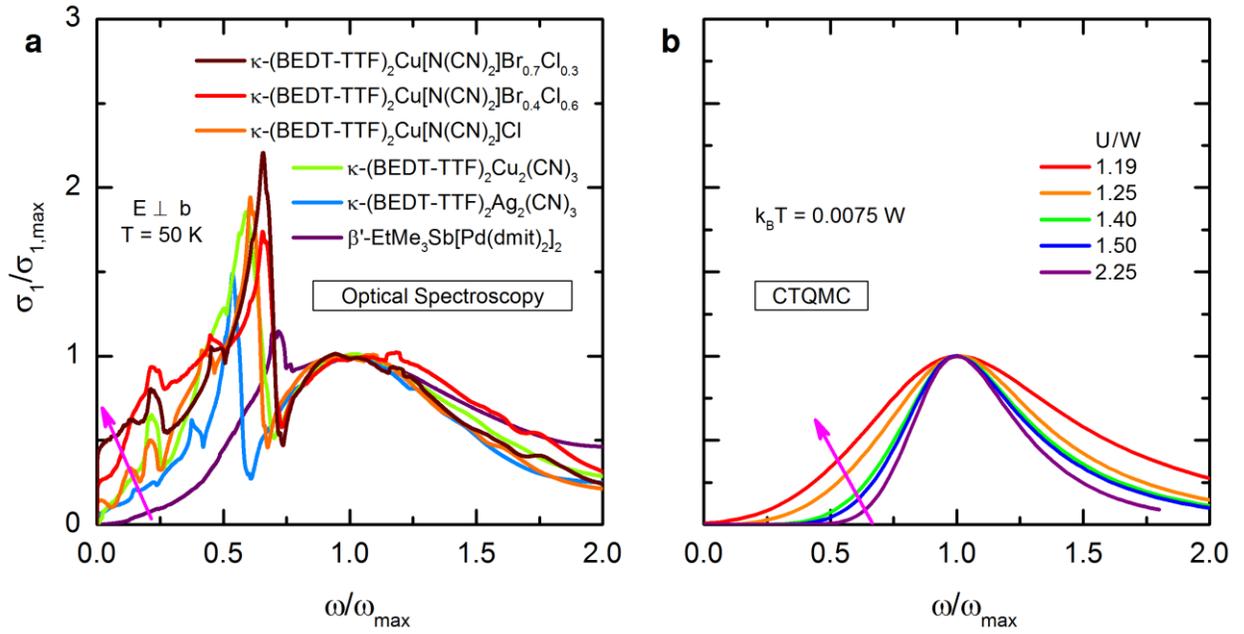

**Figure S10. Similar bandwidth-tuning effect on low-frequency absorption in experiment and theory.**

**a**, Normalized optical conductivity of two-dimensional Mott insulators with different effective correlations $U/W$: the data are smoothed and normalized to the maximum conductivity $\sigma_{max}$ of the Mott-Hubbard band and the corresponding frequency $\omega_{max}$. The $T = 50$ K data for $\beta'$-EtMe$_3$Sb[Pd(dmit)$_2$]$_2$, $\kappa$-(BEDT-TTF)$_2$Ag$_2$(CN)$_3$ and $\kappa$-(BEDT-TTF)$_2$Cu$_2$(CN)$_3$ are complemented by spectra of $\kappa$-(BEDT-TTF)$_2$Cu[N(CN)$_2$]Br$_x$Cl$_{1-x}$ with $x = 0$ and 0.4 above their magnetic order at $T_N = 25$ K, as well as $x = 0.7$ [S1,S3]. **b**, Bandwidth tuning affects the experimental optical conductivity in a similar way as predicted by theory based on the continuous time quantum Monte Carlo (CTQMC) method. Most important is the enhancement of the low-frequency conductivity when correlations are reduced, as indicated by the magenta arrows.

In Fig. S11a we plot the spectral weight (SW) integrated up to a cut-off frequency $\omega_i < (U - W)$ for both in-plane directions of the quantum spin liquids, cf. Figs. 4d and S5. The overall temperature evolution is the same for both polarizations, in line with the findings for the spin liquid compounds. Performed the corresponding SW analysis on the $\kappa$-Br$_x$Cl$_{1-x}$ series ($\omega_i = 700$ cm$^{-1}$) we conclude that these materials show a similar non-thermal enhancement of the in-gap absorption like CuCN. Note, that for $T > 50$ K the low-frequency optical conductivity measured by the integrated SW increases upon cooling although the dc resistivity (Fig. S9b) resembles an insulating behavior. There is only little difference between $x = 0$ and 0.4 on the one hand, as both are located in the Mott insulating region, and the compound with $x = 0.7$ on the other hand, which is presumably in the incoherent regime above $T_{MI} \approx 50$ K. We conclude that the non-thermal enhancement of the SW is caused by metallic fluctuations that appear on both sides of the quantum Widom line in the vicinity of the metallic phase, i.e. in the Mott state and the incoherent regime, as pointed out in Fig. S9a. This precursor of coherent quasiparticles is a robust feature observed in Mott insulators with and without magnetic order, and varying degree of geometrical frustration. Also, the effect of antiferromagnetism can be clearly separated since for $x = 0$ and $x = 0.4$ the spectral weight rapidly drops below $T_N$ when the antiferromagnetic gap opens. Apparently, charge fluctuations are suppressed more strongly by AFM order than by the intrinsic Coulomb interactions of the Mott state.

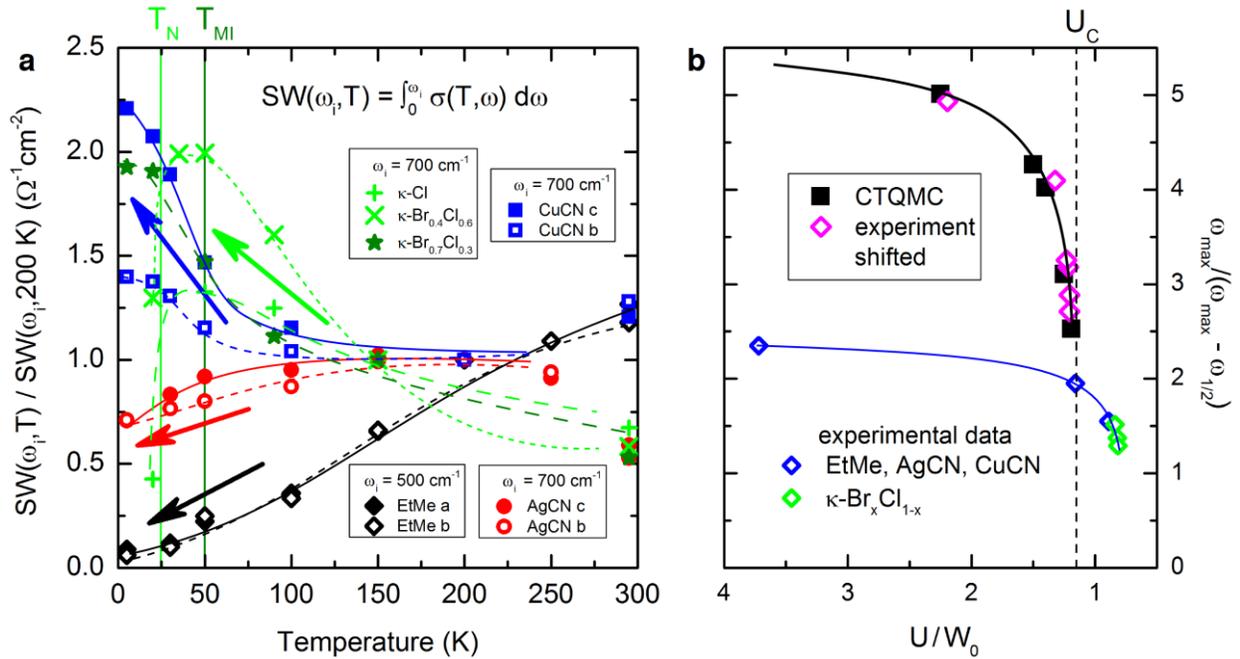

**Figure S11. Optical parameters compared between spin liquids, antiferromagnetic insulators and theory.**

**a**, Temperature evolution of the spectral weight (SW) of the three quantum spin liquid compounds $\beta'$-EtMe$_3$Sb[Pd(dmit)$_2$]$_2$, $\kappa$-(BEDT-TTF)$_2$Ag$_2$(CN)$_3$ and $\kappa$-(BEDT-TTF)$_2$ Cu$_2$(CN)$_3$. In extension to Fig. 4d, both polarizations are plotted yielding similar temperature evolution. In addition, the SW of the antiferromagnetic Mott insulators from the $\kappa$-(BEDT-TTF)$_2$ Cu[N(CN)$_2$]Br$_x$Cl$_{1-x}$ series is included [S1,S3]. Above $T_N$ these systems behave similarly as $\kappa$-(BEDT-TTF)$_2$ Cu$_2$(CN)$_3$ supporting our conclusion of the non-thermal absorption. As magnetic order sets in for $x = 0$ and $x = 0.4$, the spectral weight drastically drops due to the opening of the antiferromagnetic gap. **b**, Correlation dependence of the maximum of the optical conductivity with respect to its width of the lowest-temperature data from Figs. 2 and S5 (blue) contrasted to the $\kappa$-(BEDT-TTF)$_2$Cu[N(CN)$_2$]Br$_x$Cl$_{1-x}$ series (green) and the CTQMC results (black) from Fig. S8b. Note the reversed axis of ordinates. The discrepancy between experiment and theory is most likely due to several assumptions of the model. Still, the trends agree very well: i.e. the experimental data can be rescaled to match exactly with theory (magenta symbols). Hence, we conclude that the underlying Mott physics is reproduced reasonably well by the CTQMC calculations.

**S7 Experiment vs. Theory**

Finally, we compare the band shape of the experimentally determined optical conductivity with the spectra calculated by the continuous time quantum Monte Carlo (CTQMC) method. Fig. S11b shows the ratio of the maximum position $\omega_{max}$ of the charge-transfer band and the half-width on its low-frequency wing $\omega_{max} - \omega_{1/2}$, i.e. $U/W$ as determined in Fig. 2, with respect to the degree of electronic correlations $U/W_0$. For the experiment, we plot the lowest temperature values of Fig. 4d versus the theoretically calculated (DFT, extended Hückel) $U/W_0$ values from Table S1. The $\kappa$-Br$_x$Cl$_{1-x}$ data were extracted similarly as for the quantum spin liquids. At first glance, there is a quantitative discrepancy between experimental and theoretical data since, apparently, the theoretical bands are narrower and the calculations imply the onset of Mott insulating behavior at a larger $U_C/W_0$. However, there are many assumptions of the theoretical model that may cause deviations of the absolute values, such as the Bethe lattice instead of a triangular lattice, the dimer approximation, electron-lattice coupling, screening etc. Therefore, it is reasonable to shift the experimental curve to a different onset $U_C/W_0$ and rescale the absolute values. As the result of such a transformation, the experimental data match perfectly with the theoretical ones and their curvature compares very well. Hence, our theoretical model that only includes the charge dynamics of the Mott insulator qualitatively reproduces the electrodynamic response of our optical spectroscopy results.

## S8 Comparison with Transition-Metal Oxides

In Fig. S12a, we illustrate the experimentally covered ranges in the unified phase diagram shown in Fig. 5, including our own experiments on the three frustrated Mott insulators (optics and pressure-dependent dc transport) as well as the data reported in Refs. [S15,S13]. For comparison, we add the data of $(V_{1-x}Cr_x)_2O_3$ which exhibits a transition from a paramagnetic metal to a Mott insulator upon chromium substitution. In Fig. S12b the pressure and temperature dependence of the phase transition is plotted as obtained from dc transport measurements at variable pressure [S23]. From optical experiments [S24,S25,S26] it can be estimated that the bandwidth is approximately higher by a factor of 25 compared to EtMe; the Coulomb repulsion $U$ is around 2.6 eV. Performing the proper scaling, we can match the phase boundary of vanadium sesquioxide with the universal QWL, covering a significantly smaller region in the phase diagram. Note that $T = 300$ K on the EtMe temperature scale correspond to several $10^3$ or even $10^4$ K for vanadium oxides, cuprates or similar materials, which is well above their melting points. Thus, the small bandwidth of the organic compounds allows us to reach a temperature regime far beyond the grasp of transition metal oxides, giving unique insight into incoherent conduction processes outside the realm of quasiparticles.

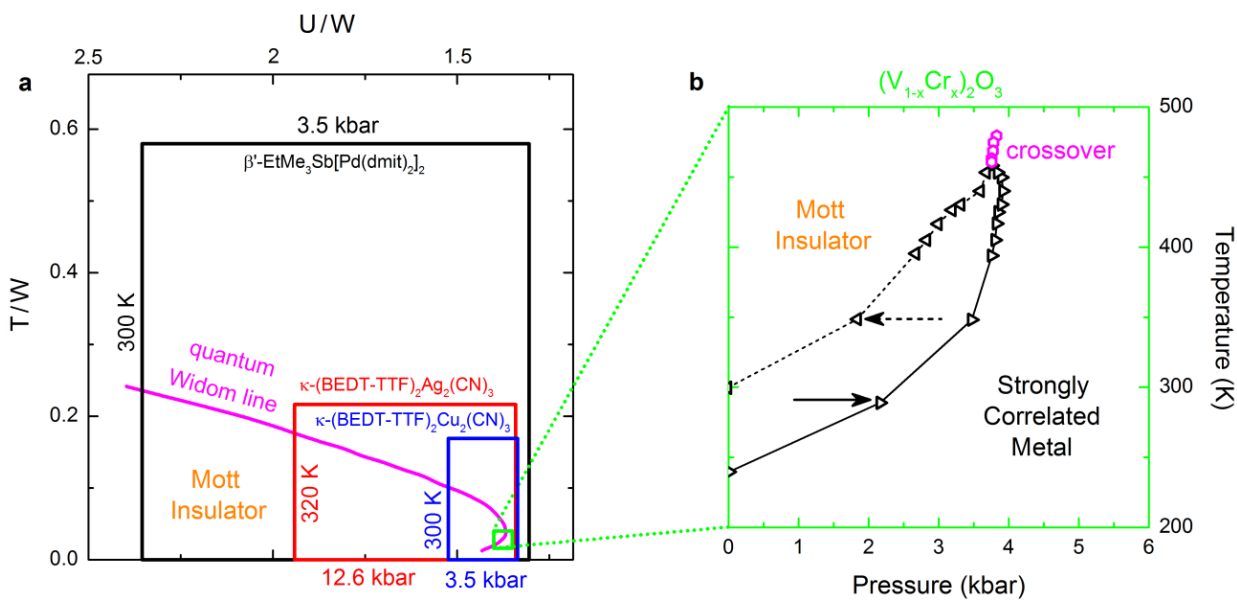

**Figure S12. Range of experimental exploration in the unified phase diagram.**
**a**, Experiments on various compounds cover the unified phase diagram of Fig. 5 to a different degree. We illustrate the respective $T/W$ and $U/W$ ranges by colored frames (cf. Section S2). As the bandwidth increases in the order $\beta'$-EtMe$_3$Sb[Pd(dmit)$_2$]$_2$ (black) - $\kappa$-(BEDT-TTF)$_2$Ag$_2$(CN)$_3$ (red) - $\kappa$-(BEDT-TTF)$_2$Cu$_2$(CN)$_3$ (blue), the corresponding range is successively reduced, both vertically and horizontally. **b**, Experimentally accessible phase boundary of $(V_{1-x}Cr_x)_2O_3$ as a function of pressure and temperature (reproduced from Ref. [S23]). By including these data in panel **a** by the green frame, it becomes obvious that a much smaller region is covered despite the larger temperature and pressure scales. This is a direct consequence of the larger energy scales, $U$ and $W$, in transition-metal oxides.